\documentclass[12pt,preprint]{aastex}
\usepackage{longtable}

\slugcomment{To appear in Astrophys. Journal}

\shorttitle{Quantitative Morphology of Galaxies }
\shortauthors{C. M. Guti\'errez et al.}

\begin{document}
\title{Quantitative Morphology of Galaxies in the Core of the Coma Cluster}
\author{Carlos M. Guti\'errez, Ignacio Trujillo\altaffilmark{1}, Jose A. L. Aguerri} \affil{Instituto de Astrof\'{\i}sica de Canarias, 
  E-38205, La Laguna, Tenerife, Spain}
\author{Alister W. Graham}\affil{Department of Astronomy, University of 
  Florida, Gainesville, Florida, USA} 
\and \author{Nicola Caon} \affil{Instituto de Astrof\'{\i}sica de
  Canarias, E-38205, La Laguna, Tenerife, Spain}
\altaffiltext{1}{Present address: Max--Planck--Institut f\"ur Astronomie, 
  K\"onigstuhl 17, D--69117, Heidelberg, Germany}
\begin{abstract} 
We present a quantitative morphological analysis of 187 galaxies in a region 
covering the central 0.28 square degrees of the Coma cluster. Structural 
parameters from the best-fitting S\'ersic $r^{1/n}$ bulge plus, where 
appropriate,    exponential disc model, are tabulated here. This sample is
complete down to a magnitude of $R$=17 mag. By examining the Edwards 
et al.\ (2002) compilation of galaxy redshifts in the direction of Coma, we 
find that 163 of the 187 galaxies are Coma 
cluster members, and the rest are foreground and background objects.
For the Coma cluster members, we have studied differences in the
structural and kinematic properties between early- and  late-type galaxies, 
and between the dwarf and giant galaxies. Analysis of the elliptical galaxies reveals 
correlations among the structural parameters similar to those previously 
found in the Virgo and Fornax clusters. 
Comparing the structural properties
of the Coma cluster disc galaxies with disc galaxies in the field, we find 
evidence for an environmental dependence: the scale lengths of the disc 
galaxies in Coma are 30\% smaller. 
A kinematical analysis shows 
marginal differences between the velocity distributions of ellipticals with 
S\'ersic index $n<2$ (dwarfs) and those with $n>2$ (giants); the dwarf galaxies
having a greater (cluster) velocity dispersion. Finally, our analysis of all 421 
background galaxies in the catalog of Edwards et al.\ reveals a non-uniform distribution in redshift with 
contrasts in density $\sim 3$, characterized by a void extending from 
$\sim 10,000$ to $\sim 20,000$ km s$^{-1}$, and two dense and extended 
structures centred at $\sim 20,000$ and $\sim 47,000$ km s$^{-1}$.

\end{abstract}


\section{Introduction}

The properties of galaxies can vary depending on whether they reside in 
dense galaxy clusters or the field. The most remarkable example of 
this is the morphology--density relation (Dressler 1980) in which 
the proportion of elliptical galaxies increases toward the cores
of rich clusters. The morphology of galaxies in clusters has been based, mostly,
on a visual classification scheme. However, visual
classification is only the first step in the characterization and description 
of galaxies.  It is necessary to conduct a quantitative morphological
analysis of galaxies in clusters to answer basic questions like:
% Are the sizes of the elliptical galaxies in clusters different than their
% field counterparts?, or, 
Are the properties of spiral galaxy discs, such as their scale-lengths, 
affected by the enviroment? 
Such a study is also required to make a detailed comparison
with, and therefore test, current theoretical predictions 
(e.g.\ Moore et al.\ 1999; Gnedin 2003). 

The proximity and richness of the Coma cluster has made it one of the most 
studied galaxy clusters. Since Godwin, Metcalfe, \& Peach (1983) published 
the first wide-field galaxy catalog using photographic photometry, many other surveys have been
conducted  both in the central parts of this cluster (e.g., Jorgensen \&
Franx 1994;  Karachentsev et al.\ 1995; Bernstein et al.\ 1996; Lobo et al.\
1997; Secker  \& Harris 1997; Trentham 1998) and covering larger areas
(e.g., Kashikawa et al.\  1995; Terlevich, Caldwell \& Bower  2001; Beijersbergen
et al.\ 2002).  A recent survey combining wide field photometry and
spectroscopy has been presented in Komiyama et al.\ (2002) and Mobasher et
al.\ (2001). There are numerous morphological studies of galaxies within the
Coma cluster, both in the optical (e.g., Rood \& Baum 1967; Dressler
1980; Lucey  et al.\ 1991; Jorgensen \& Franx 1994;  Andreon et al.\ 1996;
Andreon, Davoust, \& Poulain 1997; Gerbal et al.\ 1997; Kashikawa et al.\
1998; Mehlert et al.\ 2000; and Komiyama et al.\ 2002) and in the near-infrared 
(e.g., Pahre 1999; Mobasher et al.\ 1999; Khosroshahi et al.\ 2000).

In this paper we present the morphology and structural parameters of
galaxies in the central region of the Coma cluster (0.28 square
degrees).  We wish to stress that our analysis uses for the first time
velocity data to establish cluster membership.  
Furthermore, and importantly, we do not a priori assume to know what the 
distribution of light is in elliptical galaxies or the bulges of spiral 
galaxies.  That is, rather than force the $r^{1/4}$ model on these 
systems, we use S\'ersic's (1968) $r^{1/n}$ model in an effort to 
{\it measure} the distribution/concentration of light. 
A detailed 
analysis of the relation between galaxy light concentration and galaxy
environment was addressed in a previous paper (Trujillo et al.\ 2002a,
hereafter T02A).  One of our present objectives is to study the various 
correlations among the structural parameters, and to search for possible 
differences according to
morphological type or local conditions within the cluster.  A study of
a rich and nearby cluster like Coma is also very useful for
establishing a local reference for studies of clusters at intermediate and 
high redshifts.

Section~2 describes the observations and the compilation of redshifts. The 
method to determine the quantitative morphology of galaxies is outlined in 
Section~3. In Section~4 we explore the relationships between the structural 
parameters and also with the environment. Section~5 summarizes the main 
results of the paper.

\section{The sample and observations}

\begin{figure}  
\plotone{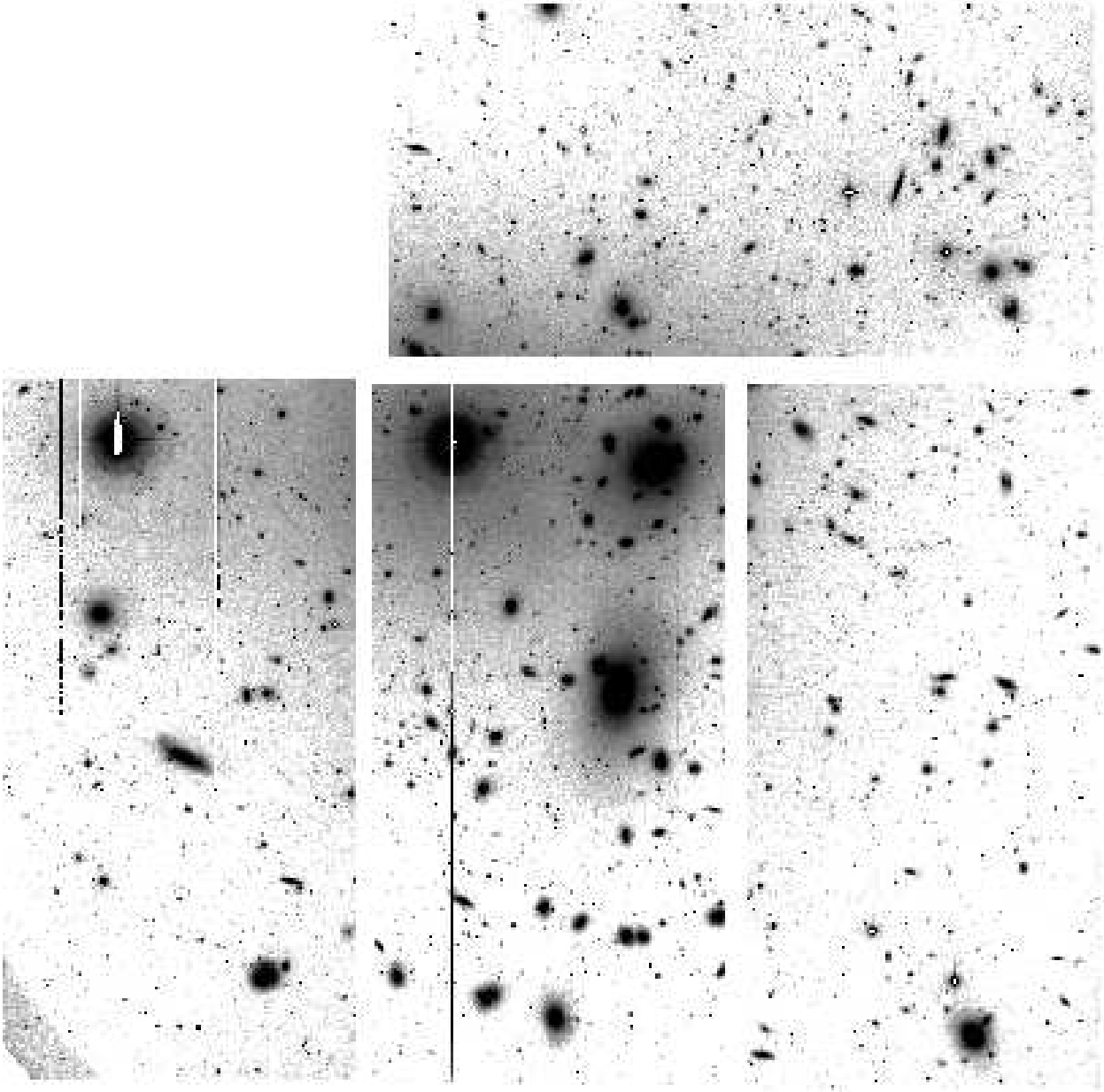} 
\protect\caption[]{Gray-scale image of the Coma cluster in the Sloan $r$ 
filter. The image was obtained with the Wide Field Camera at the Isaac Newton 
Telescope, and is the result of the co-addition of 13 individual exposures 
giving a total exposure time of 3900 seconds. The angular size is 
$\sim 34^\prime\times 34^\prime$. North is to the left and East to the bottom.
\label{wfc}}
\end{figure}

We have performed a quantitative morphological analysis of galaxies in the
Coma cluster using a deep image taken in April 2000 with the Wide Field Camera
(WFC) at the 2.5 m Isaac Newton Telescope (INT) at the Roque de los Muchachos
Observatory. The observations are described in detail in Mar\'\i n-Franch \&
Aparicio (2002). Here, we outline the more relevant facts. The image was taken
through a Sloan $r$-band filter, with a total exposure time of 3900 s
(13$\times$300 s). The observations covered the inner 0.28 square degrees of 
the central part of the cluster (see Fig.~\ref{wfc}). Conditions were 
photometric. 
We performed a standard data reduction, comprising subtraction of bias, flat
field corrections and co-addition of individual exposures. Each chip was
calibrated using standard Landolt stars (Landolt 1992). The seeing in the
final (combined) image, measured using bright, unsaturated stars, was 1.1 
arcsec.
The limiting magnitude for these observations was $r\sim$ 23.5 mag. The four 
frames in Fig.~\ref{wfc} correspond to the four (2k $\times$ 4k) CCDs 
(with a scale of 0.333 arcsecconds pixel$^{-1}$) of the 
WFC. The position of the CCDs in the Camera produces gaps of $\sim 16$ to
$\sim 27$ arcseconds on the sky.

The $B$ magnitudes and $B-R$ colors given in Table~1 (see next Section), were 
taken from the catalog of Godwin et al.\ (1983). The recessional velocities are 
from the compilation by Edwards et al.\  (2002). These catalogs cover 
a region of 2.65 square degrees centred on the Coma cluster and largely 
overlap our observed region. 
The compilation by Edwards et al.\ comprises 1174 galaxy redshifts of Coma 
members, and foreground and background objects.
The mean velocity of the Coma cluster is $\sim 7,000$ km s$^{-1}$, and its velocity 
dispersion is $\sim 1,000$ km~s$^{-1}$. Our criterion for membership is the 
same as the one used by Edwards et al.: Coma members are those objects with 
velocities in the range $4,000 \le cz \le 10,000$ km s$^{-1}$.  This range 
corresponds to a $\sim 3\sigma$ cut on the velocity distribution of the Coma 
cluster galaxies.
With this criterion, the number of Coma members in the above velocity catalog 
is 745. 
The sample 
analyzed in this paper has been selected by magnitude, and includes only six 
galaxies with unknown redshifts (see next sections), so incompleteness effects 
are negligible. We select for our structural analysis (see next section) the 
187 galaxies with $R\le 17$ mag.

\section{Quantitative morphology}

To quantify 
the properties of each galactic structural component (bulge and disc)
we use a parametric model to describe the observed radial profiles. 
Elliptical galaxies and the bulges of spiral galaxies are modeled with 
a S\'ersic law; discs are described by an exponential profile. 
The effects of atmospheric
blurring on S\'ersic profiles (Trujillo et al.\ 2001b, 2001c) have been
taken into account using the algorithm described in Trujillo et
al.\ (2001a; hereafter T01A) and Aguerri \& Trujillo (2002).
In the case of optical ground based observations, the PSF is dominated by 
atmospheric blurring and can be approximated by a Gaussian or Moffat function.
For the observations presented here, we determined the PSF from the 
radial profile of bright unsaturated stars in the image; it was 
found to be well 
described by a Moffat function with $\beta=2.5$ and FWHM=$1\farcs 1$.

The S\'ersic (1968) profile can be written as 
\begin{equation}  
I(r)=I_b(0)\exp[{-b_n(r/r_{\rm e})^{1/n}}]. 
\end{equation}
This law is a generalization of the de Vaucouleurs (1959) profile and has been
used widely in recent years as a good description of the profiles of dwarf 
and giant ellipticals and the bulges of spiral galaxies (see, for example,  
Caon, Capaccioli, \& D'Onofrio 1993; Andredakis, Peletier, \& Balcells 1995; 
Graham \& Colless 1997; 
Balcells et al.\ 2003). The parameters of the model are the central intensity 
$I_b(0)$, the effective radius $r_{\rm e}$, and the S\'ersic index $n$. The 
quantity $b_n$ is defined so that the effective radius encloses half of the 
total luminosity (i.e., $b_n$ is the solution of 
$\Gamma (2n)=2\gamma(2n,b_n)$, where $\Gamma $ and $\gamma$ are the Gamma 
function and the incomplete Gamma function respectively). The relation between 
the central intensity and the intensity at the effective radius is given by 
$I(r_{\rm e})=I_b(0)\exp(-b_n)$. Discs were modeled with an exponential 
profile, such that 
\begin{equation}  
I(r)=I_{\rm d}(0)\exp (-r/h),
\end{equation}
where the parameters are the scale length $h$ and the central intensity 
$I_{\rm d}(0)$.

We assume projected elliptical symmetry for the bulge and disc, with the 
ellipticities ($\epsilon_{\rm b}$ and $\epsilon_{\rm d}$ respectively), 
in general, different for each component. The observed 
ellipticity at small radii is of course smaller than the true ellipticity because the 
seeing tends to make the isophotes rounder. This effect is particularly 
important for the numerous dwarf galaxies and spiral galaxy bulges with 
effective radii as small as 2-3 arcseconds. Our algorithm takes this effect 
into account, and the intrinsic ellipticities for the bulge and the disc are
determined simultaneously with the other structural parameters.

We tried to use the minimum number of components (i.e. parameters), so we 
proceeded as follows.  Every galaxy's major-axis surface brightness {\it and} 
ellipticity profile, generated using the \textsc{iraf} task \texttt{ELLIPSE}, 
were simultaneously fitted with a two component model (bulge plus disc).  
The total flux of both components is computed by using the S\'ersic analytical 
expressions extrapolated to infinity.  For those 
objects in which we obtained a bulge-to-total luminosity ratio $B/T>0.6$, we 
checked if it was possible to obtain a good fit (i.e., with a reduced
chi-square value as small or smaller than in the two component case) with just
a bulge component. When that was the case, we considered the object to 
be an elliptical galaxy 

Using Monte Carlo simulations we have determined the uncertainty in the
computed parameters. Details of how these simulations are constructed can be
found in T01A and in Aguerri \& Trujillo (2002). We created 150 artificial 
galaxies with structural parameters randomly distributed in the following 
intervals:
\begin{itemize}

\item  bulge--only structures: 13 mag $\leq R \leq 19$ mag, 1$'' \leq r_{\rm e}
\leq 10''$, $0.5 \leq n \leq 4$, and $0\leq \epsilon \leq 0.6$ (the lower limit
of $n=0.5$ is due to the physical restrictions pointed out by Trujillo et al.\
2001b).

\item  bulge+disc \ structures: $13$ mag $\leq R \leq 19$ mag, $1'' \leq 
r_{\rm e}
\leq 10''$, $0.5\leq n \leq 4$, $0\leq \epsilon_{\rm b} \leq 0.4$, $5'' 
\leq
h \leq 25 ''$, $0\leq B/T \leq 1$, and $0\leq \epsilon_{\rm d} \leq 0.6$.
\end{itemize}

A variety of starting parameters were used to ensure that our fits did not get
trapped in local $\chi^2$ minima. From these simulations we find that the bulge
and disc parameters (we include here the intrinsic ellipticity of each
component) can be  determined with an accuracy of $\sim 10$\% for
galaxies with $R\le 17$. Our structural analysis was therefore carried out only on
the 187 galaxies brighter  than $R = 17$ mag; the results are presented in
Table~1. Of these 187 galaxies, 163 have recessional velocities between 4,000
and  10,000 km s$^{-1}$; for 32 of these, the parameters could not be measured 
with sufficient accuracy for various reasons, such as proximity to bright 
stars, uncertain deblending, bad columns of the detector, etc.  These galaxies
are flagged  with a negative sign preceding their $B/T$ ratio.  This group also
includes galaxies which are irregular or peculiar.  Also, because the
structure of cD galaxies (NGC~4874 and NGC~4889) is poorly  understood, these
galaxies will also be excluded.  The analysis presented in the following sections
will be restricted to the 129  regular galaxies belonging to Coma and having
reliable parameters. 

\begin{figure}  
\plotone{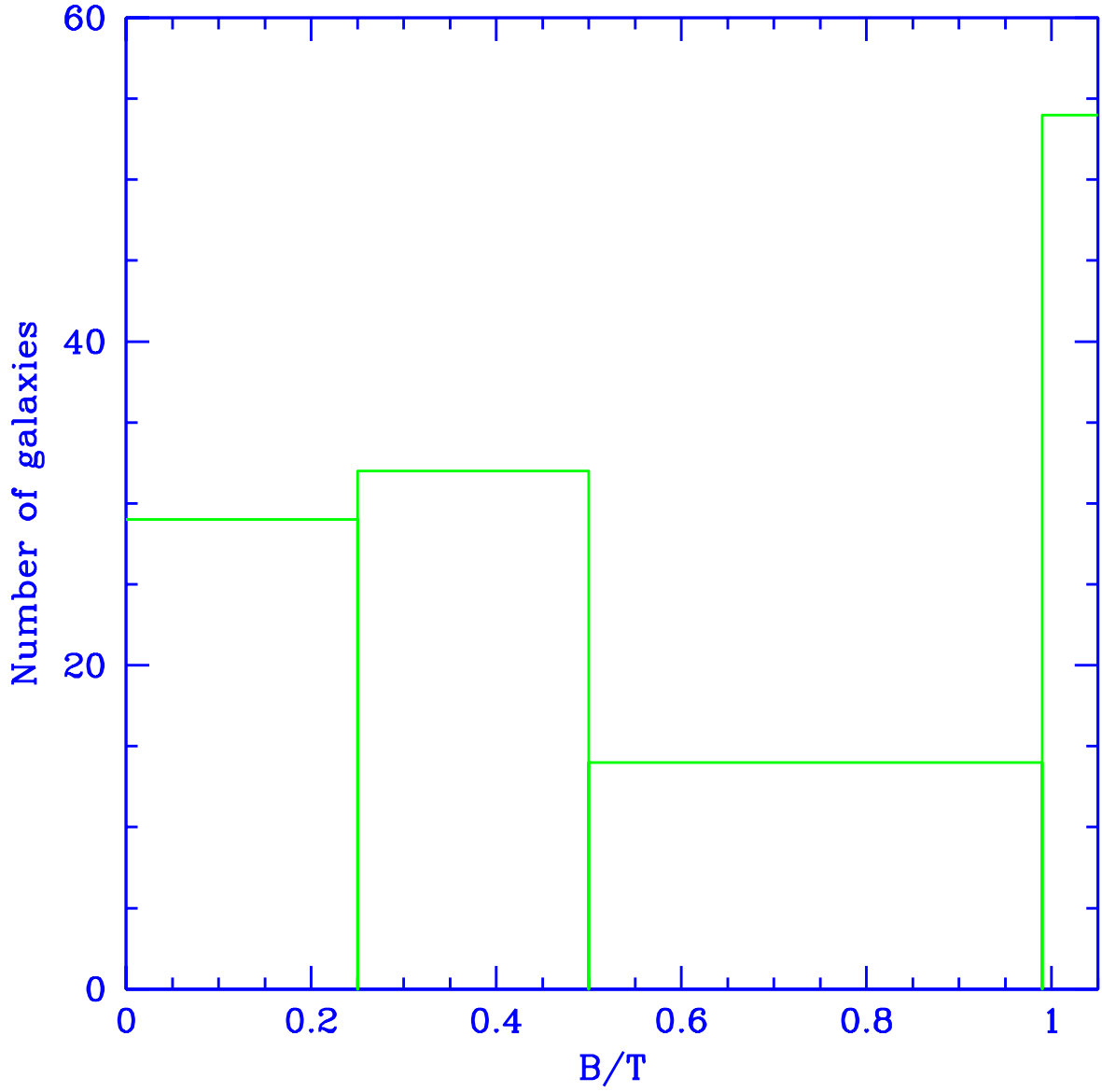}
\caption{Histogram showing the number of Coma galaxies analysed in 
this paper as a function of the bulge-to-total luminosity ratio.
\label{fig_bt2}}
\end{figure}

\begin{figure}  
\plotone{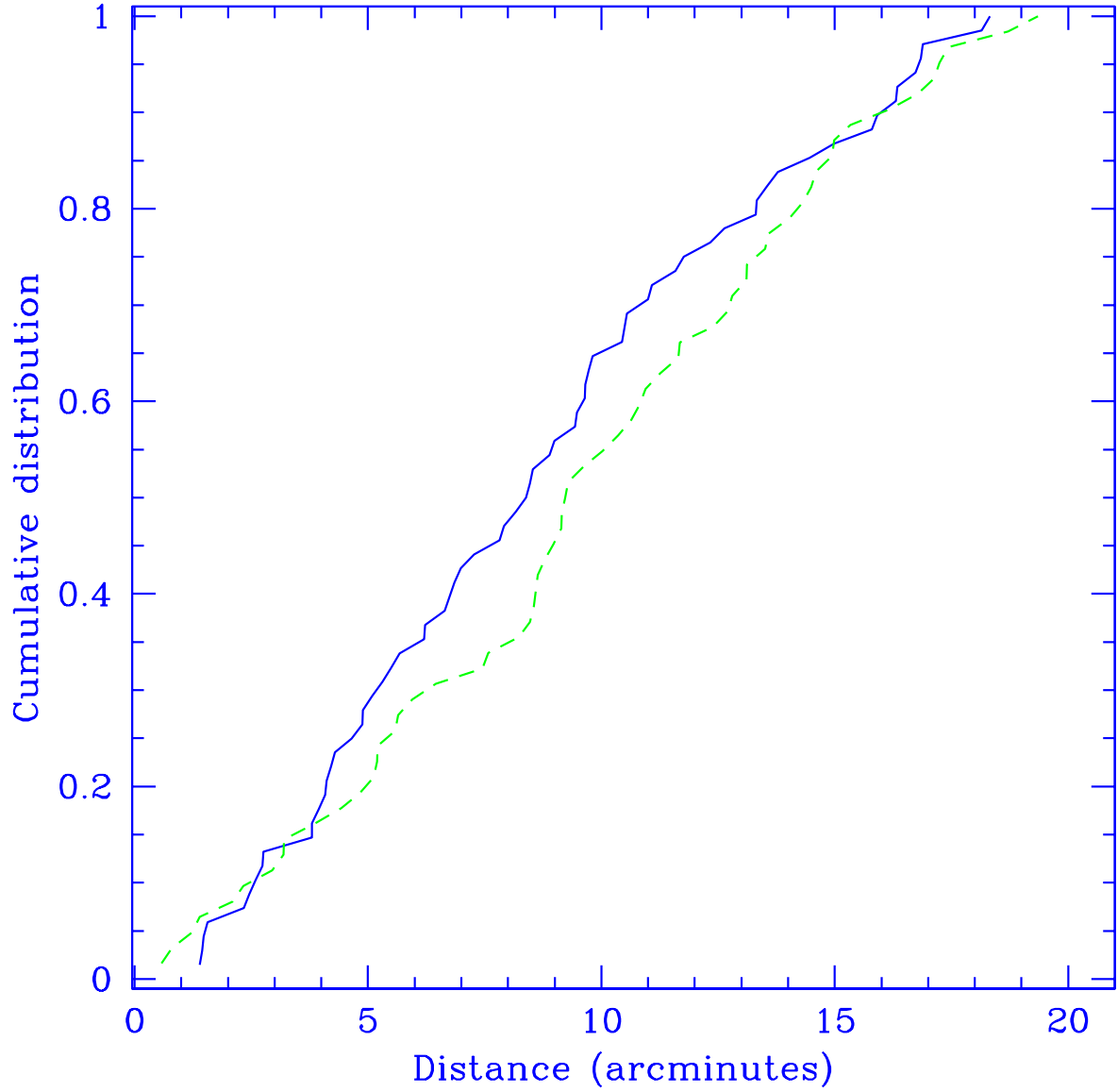}
\caption{Cumulative probability distribution of early
($B/T>0.5$) ({\it solid line}) and late ($B/T\le 0.5$) ({\it dashed line}) 
type galaxies as a function of the distance to the center of the cluster.
\label{fig_dist_bt}}
\end{figure}

Following the notation and taxonomy discussed in T01A, we have differentiated 
the galaxies according to the bulge-to-total luminosity ratio ($B/T$). We will 
consider the value $B/T=0.5$ as the separation between early (E/S0) and late 
(S) type galaxies.  According to this criterion, 61 galaxies were classified 
as late-type, and 68 galaxies as early-type. In the early-type group, 14 
objects show evidence of discs, while 54 are pure ellipticals.  In the text we
will often refer to  objects that have been modeled with  only a bulge
component as elliptical galaxies. This does not necesarily imply  some specific
internal kinematics for these objects.  Figure~\ref{fig_bt2} shows a histogram
of the $B/T$ values.  Figure~\ref{fig_dist_bt} shows the cumulative
distribution of early- and late-types as a function of the distance to the
cluster center. Although the figure  seems to indicate some evidence for the 
early-type galaxies to be distributed closer to the  center than late-types, a
Kolmogorov-Smirnov test shows that this is not statistically significant
(similar results were obtained considering the value $B/T=0.3$ as the
separation between early- and late-type galaxies). A more detailed analysis of
this  point is presented below. 

In Figure~\ref{gmp} we compare the $R$-band magnitudes quoted in the GMP catalog
with the galaxy magnitudes derived from our (Sloan-$r$ passband) structural
parameters.  Although the agreement is very good, there is a $\sim 0.2$ mag
zero-point offset such that our estimates are fainter.   This may be due to an
offset in the relative calibration or to the existence of a small color term
between sloan-$r$ and the $R$-band magnitudes. There is, however, no significant
magnitude-dependent bias in our estimation of the magnitudes.

\subsection{Comparison with previous morphological studies of Coma} 

Although many studies in the past have been devoted to a morphological 
analysis of galaxies in the Coma cluster (see references in Section 1), the 
comparison between the structural parameters derived for individual galaxies 
from different papers is not straightforward and is often only of a limited nature.  
Some of the previous studies provide no tables of parameters but
present only statistical results (e.g., Gerbal et al.\ 1997), or the analysis 
is purely qualitative, or they derive the structural parameters in a model 
independent way (e.g.\ Andreon 1996).  In our sample we 
have 54 galaxies in common with Dressler (1980). If we compare the 
morphological types derived there with those inferred from our $B/T$ ratio, 
we find good agreement. In a few cases, we classified objects as S0 or spiral galaxies
that were classified as pure E by Dressler.  
This is likely explained as a consequence of the different sensitivities of 
the two studies (the faint discs we found in some of our objects were not 
detected by Dressler). 
For only four galaxies classified by Dressler as S0, and as pure E by us, can 
the results be considered discrepant.  However, this level of discrepancy 
($\sim 8$\% ) is in agreement with the differences expected among 
classifications by different authors (see Lahav et al.\ 1995; Andreon \& 
Davoust 1997).

Graham \& Guzm\'an (2003) have analysed a sample of 15 dwarf elliptical galaxies
in the Coma cluster imaged with the {\itshape Hubble Space Telescope},
finding that all but two of them have a central point-like component. 
In principle, ignoring the presence of this central component could affect the 
determination of the structural parameters of bulges and discs (Balcells et 
al.\ 2003). We have checked whether this could be the case for our sample. 
We have seven objects in common with Graham \& Guzm\'an (2003). 
Although in our model we have not accounted for a central source, 
and we used major-axis light-profiles while Graham \& Guzm\'an 
used geometric-mean light-profiles, the 
parameters obtained for the galaxies in common are in good agreement. 
For instance, the typical deviation between  estimates of the S\'ersic 
indices is $\Delta n \le 0.25$. Only in one case (GMP 3292) the two 
analyses give significantly different values. Therefore, at least for the dwarf
ellipticals in Coma, the presence of nuclear components may not have a large 
effect on our estimation of their structural parameters.

\section{Results and Discussion}

\subsection{Early-type galaxies} 

Caon et al.\ (1993) and D'Onofrio, Capaccioli, \& Caon (1994) reported the 
existence of a correlation between the  S\'ersic index $n$ and the 
{\it model-independent} total luminosity of elliptical galaxiess in the Virgo and Fornax 
clusters. This correlation has been shown to hold also for the bulges of 
spirals (Andredakis, Peletier, \& Balcells 1995; Graham 2001; 
Balcells et al.\ 2003; MacArthur, Courteau, \& Holtzman\ 2003), and also 
extends to the dwarf elliptical regime (Young \& Currie 1994, 1995; Binggeli \&
Jerjen 1998, Graham \& Guzm\'an 2003).  The correlation is such that more
luminous bulges tend to have larger values of $n$, their light distributions are 
more centrally concentrated. . 
In this section we analyse this and other possible correlations existing 
among the structural parameters of the galaxies in our sample, and compare 
the results with previous studies. The study  by Caon et al.\ was
conducted in the $B$-band; to make a proper comparison we have converted our surface 
brightness values obtained in the $R$-band using the $B-R$ color given in Table 1. 

\begin{figure}  
\plotone{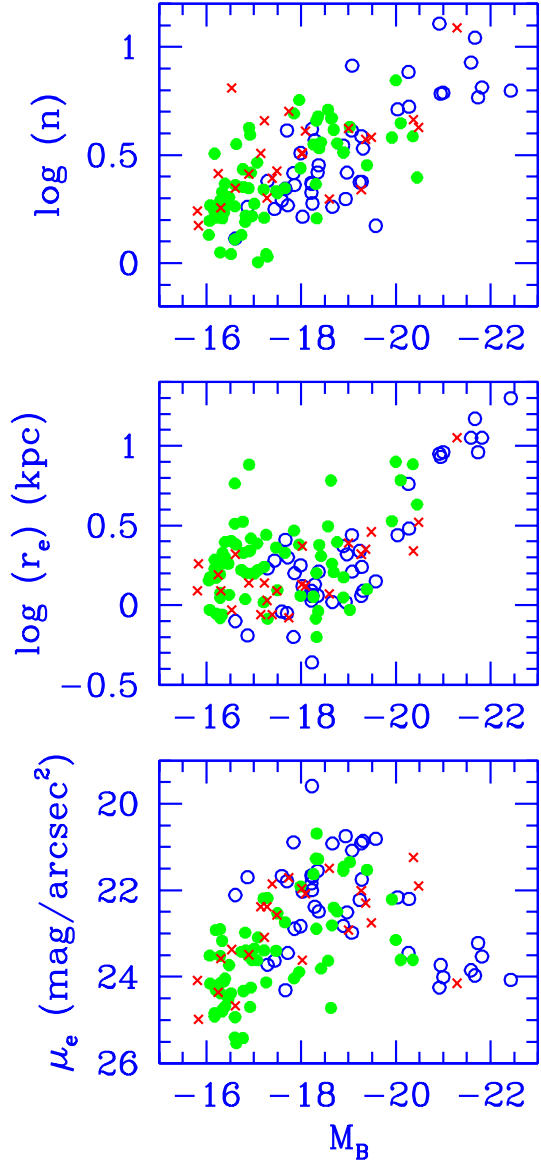}
\caption{Scaling relations between the {\it model-independent} $B$ band luminosity and the 
structural parameters of bulges in the Sloan $r$ band. From top to bottom in 
the vertical axis is represented the logarithm of the S\'ersic index, $n$, 
the effective radius, $r_{\rm e}$, and the surface brightness $\mu_{\rm e}$. 
Filled circles represent the early-type (E + S0) galaxies of Coma analysed in 
this paper, while open circles and crosses represent early-type galaxies in 
Virgo and Fornax respectively.
\label{fig_param1}}
\end{figure}

Figure~\ref{fig_param1} shows the relation between the {\it model-independent}
$B$-band magnitudes from Godwin et al.\ (1983; we assume $h\equiv 
H_0/100=0.7$), and the three S\'ersic 
parameters we obtained for the early-type galaxies in our sample. We also plot the
data for the galaxies analysed by Caon et al.\ and D'Onofrio et al.\  in the 
Virgo and Fornax clusters. The distributions of the structural parameters of 
elliptical galaxies in the three clusters are similar, although there are 
more low-luminosity dwarf ellipticals from Coma delineating the lower arm
of this forked distribution (see Graham \& Guzm\'an 2003). 

\begin{figure}  
\plotone{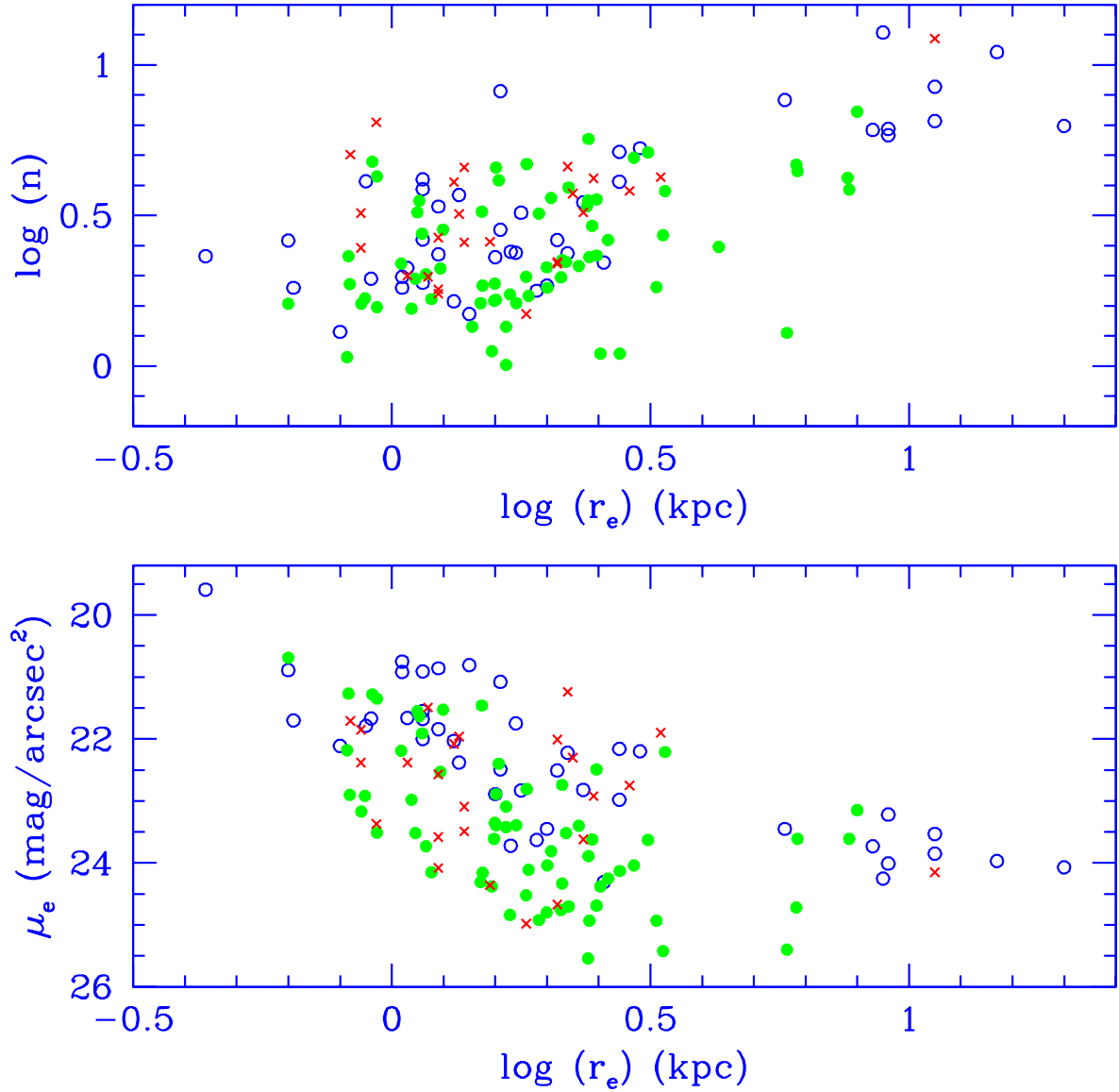}
\caption{Relation between the three structural parameters of
the early-type (E + S0) galaxies in Coma (filled circles), Virgo (open 
circles) and Fornax (crosses).
\label{fig_param2}}
\end{figure}

Another way to show the correlations between the three S\'ersic parameters is 
presented in Figure~\ref{fig_param2}. Galaxies from the three clusters exhibit 
similar values and relations between the parameters. The correlation between the 
S\'ersic index $n$ and luminosity (Figure~\ref{fig_param1}) is clear and holds 
for both the giants and dwarfs. 
The rough correlation found between $n$ and $r_e$ is similar to the one found
by Caon et al.\ (1993) (see also Young \& Currie 1995 and Graham et al.\ 1996). 
For the spiral galaxies a  similar correlation also exists between the bulge 
index $n$ and the bulge-to-total luminosity ratio (Andredakis et al.\ 1995, 
Graham et al.\ 2001; Balcells et al.\ 2003). Trujillo et al.\ (2002b) have 
interpreted this as a consequence of the relation between $n$ and the 
luminosity of the bulges. 
The main  difference between the galaxies in the three clusters
is a group of bright galaxies in Virgo which are absent in the core of Coma
and the Fornax sample. 
Objects with similar magnitudes in Coma are the two cD galaxies NGC~4874 and 
NGC~4889, which have been excluded in this analysis.

For the other two parameters (surface brightness and 
effective radius) one can distinguish two different regimes, with a transition 
region at $-20\le M_B\le -18$ close to the usual limit adopted to separate 
dwarf and ordinary ellipticals. 
Traditionally, they have been considered as two separate families of objects, 
although the exact separation in magnitude between each class 
is somewhat arbitrary. Edwards et al.\ (2002) considered dwarf galaxies as 
objects with $B\ge 18$ mag. This corresponds to M$_B\ge -16.84$ mag. 
The point here is whether the relation between structural parameters and, 
therefore, the origin of giants and dwarfs is different and justifies this 
distinction. Graham \& Guzm\'an (2003) argued that the continuity between $n$, 
and central bulge surface brightness, with luminosity demonstrates that 
dwarfs and ordinary ellipticals do not 
constitute two separate families of objects.  They explained the apparently 
different relations between $\mu_{\rm e}$ and luminosity (and $\mu_{\rm e}$ and 
$r_{\rm e}$) for the high- and low-luminosity ellipticals 
as an expected consequence of the above linear trends.

\subsection{Disc galaxies}

\begin{figure}  
\plotone{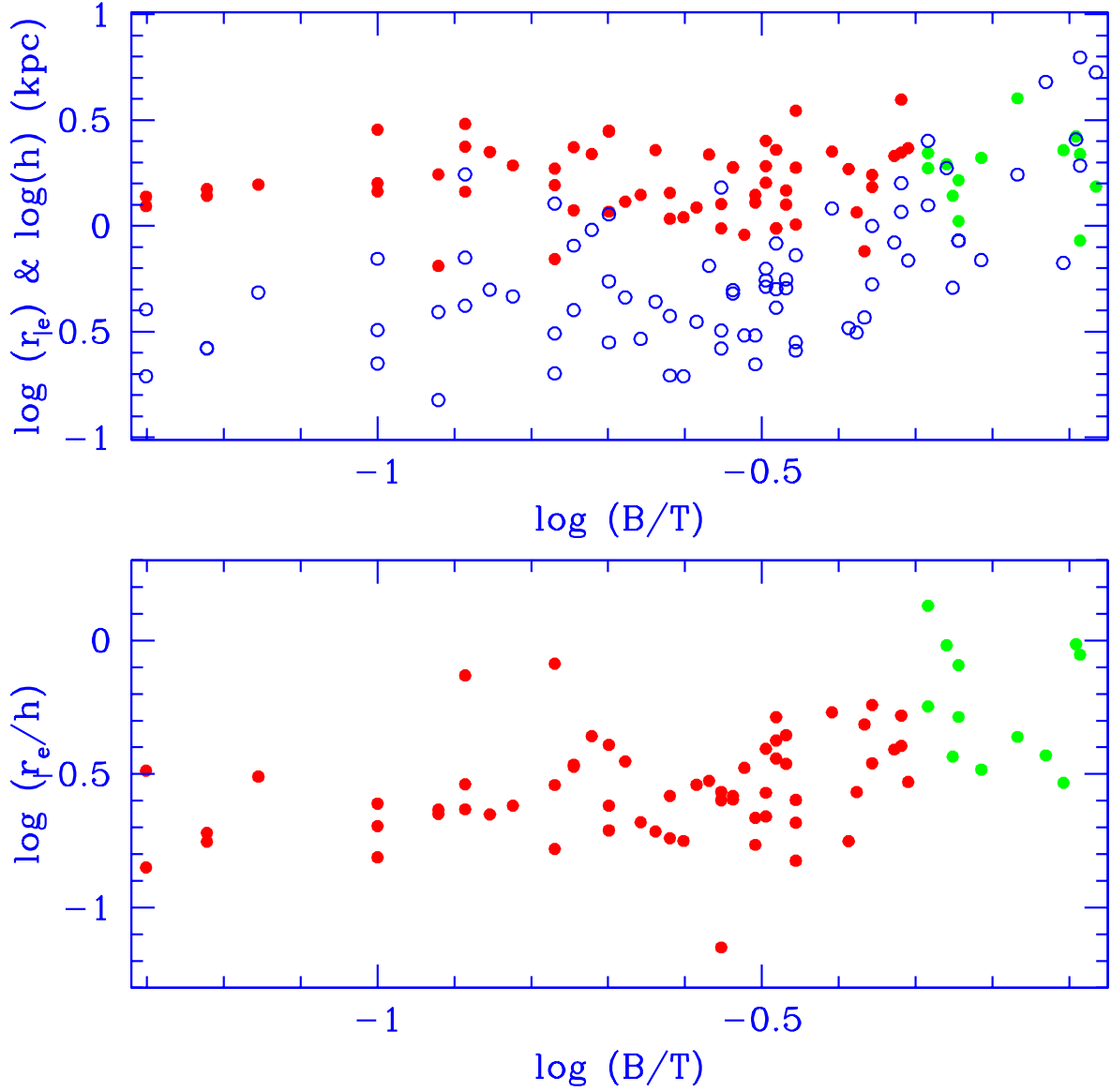}
\caption{\textit{Top}: Scale lengths of discs (filled circles) and 
bulges (open circles) as a function of $B/T$. \textit{Bottom}: disc-to-bulge 
size ratios as a function of $B/T$.
\label{fig_rerd}}
\end{figure}

We have characterized the sizes of bulges and discs using the effective 
radius $r_{\rm e}$, and scale length $h$, respectively. Figure~\ref{fig_rerd}
shows the relative sizes of bulges and discs for our sample of 14 S0 and 61 
spiral galaxies. From these plots we see how the ratio $r_{\rm e}/h$ is 
constant for galaxies with $B/T< 0.3$ (most of the spirals), with typical 
values in the range 0.15--0.30.  
For galaxies with $B/T>0.3$ this ratio increases to 0.6 and higher. 
Three early-type spiral galaxies have values $r_{\rm e}/h \ge 1$ 

\begin{figure}  
\plotone{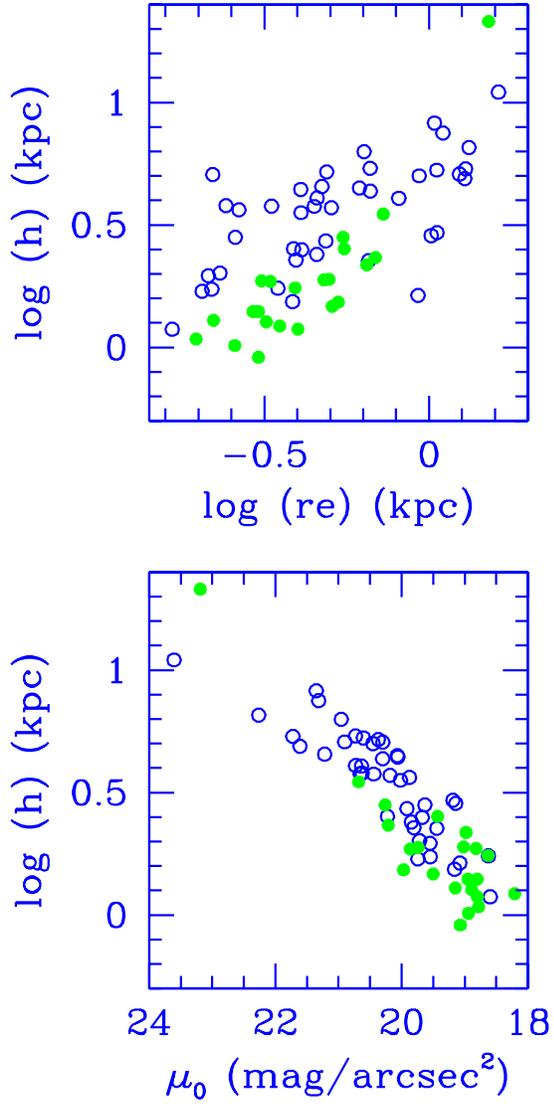}
\caption{Comparison between structural parameters of spiral galaxies
with magnitudes in the range $-22\le M_R\le -20$ in the Coma cluster (filled 
circles) and in the field (open circles).
\label{fig_disc3}}
\end{figure}

These values are somewhat larger than those obtained for spiral galaxies in the field 
(Graham 2001, 2003; MacArthur et al.\ 2003). To better understand the meaning of
these results, we have compared the properties of the discs in the field 
with those in the core of Coma.  
Disc galaxies brighter than $M_R=-22$ mag are not present in our 
sample. On the other hand, Graham's sample is not complete for galaxies 
fainter than $M_R=-20$ mag, thus we use the common range $-22$ mag $\le 
M_R\le -20$ mag. The comparison is presented in Figure~\ref{fig_disc3}.  

We obtained mean values of $r_{\rm e}/h$=0.24 for Coma and 0.17 for 
Graham's sample. A Kolmogorov-Smirnov test rejects the hypothesis that
the two $r_{\rm e}/h$ distributions are the same at $\ge 99.9$\%. 
Applying Student's t-test,  
the mean values of $r_{\rm e}/h$ in both distributions  are different at the 
99.9 \% level (assuming the two  distributions have unequal variances). The larger
average value of $r_{\rm e}/h$ for galaxies in the denser 
environment of the Coma cluster is compatible with the idea that the discs of spiral 
galaxies in the center of clusters are smaller than the discs of field 
galaxies with similar magnitudes and bulges.  In fact, if we 
assume that the sizes of the bulges are less affected by the enviroment than 
the size of the disc, we can estimate that the discs of the galaxies in the
center of the cluster have 30\% smaller scale-lengths.  Aguerri et
al.\ (2003) have found similar results from a study of the morphology of
galaxies in a larger region of the Coma cluster. 
The reduction in the sizes of the discs is in agreement with what is
expected in high density enviroments. In these enviroments tidal
forces play a crucial role truncating and heating the infalling disc
galaxies (Moore et al.\ 1999, Gnedin 2003).
This could also be explained if the suppression of star formation, 
as has been proposed by Balogh, Navarro, \& Morris (2000), is more effective 
in the outer parts of the (cluster) spiral galaxies.  As expected, due to the 
smaller values of $h$ but similar luminosities, we find that in the $h-\mu_0$ 
plane (Figure~\ref{fig_disc3}) cluster disc galaxies tend to have 
brighter central surface brightnesses.

\subsection{Ellipticities}

\begin{figure}  
\plotone{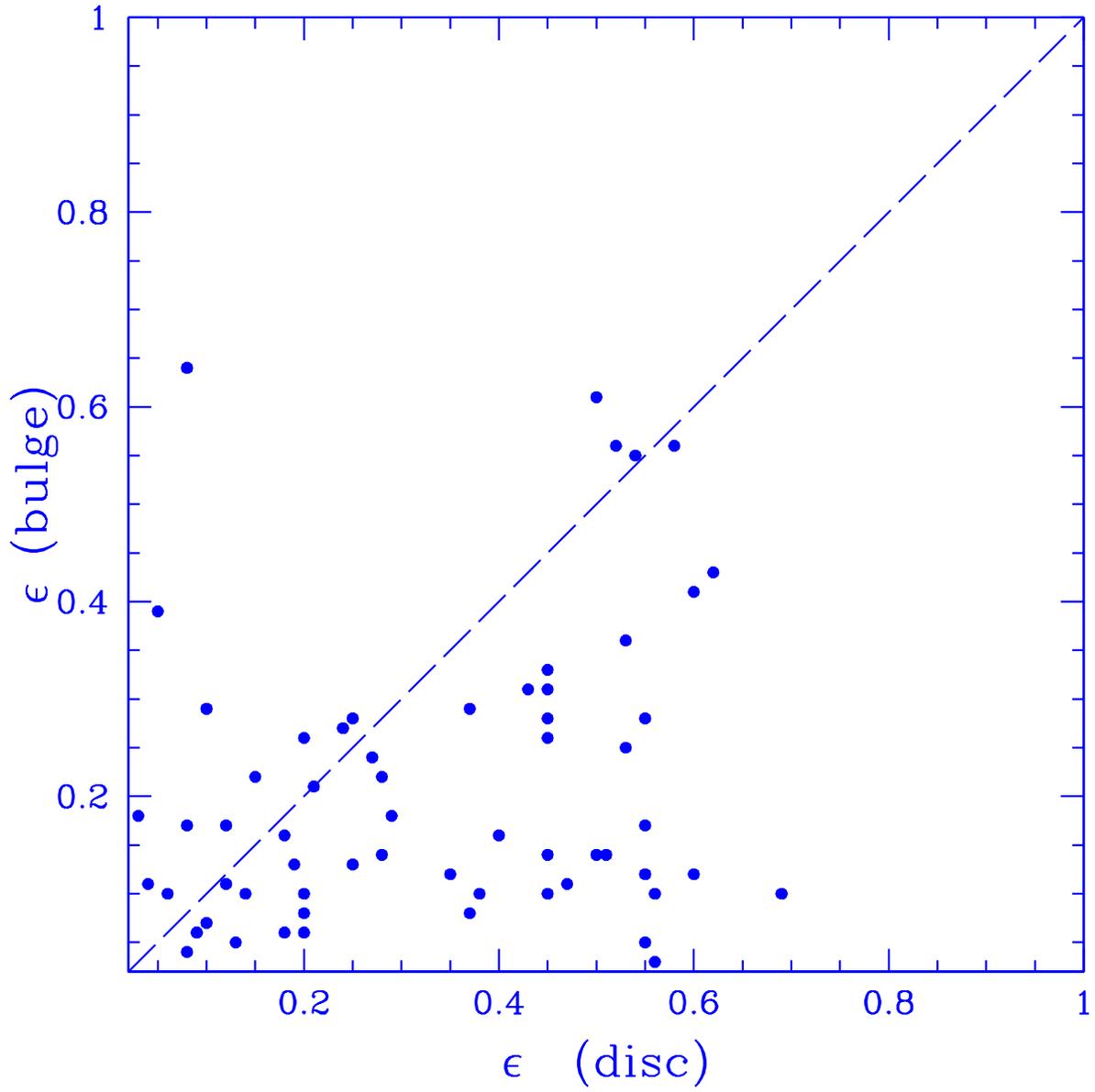}
\caption{Projected ellipticities of bulges and discs for galaxies 
with two components.
\label{fig_edeb}}
\end{figure}

\begin{figure}  
\plotone{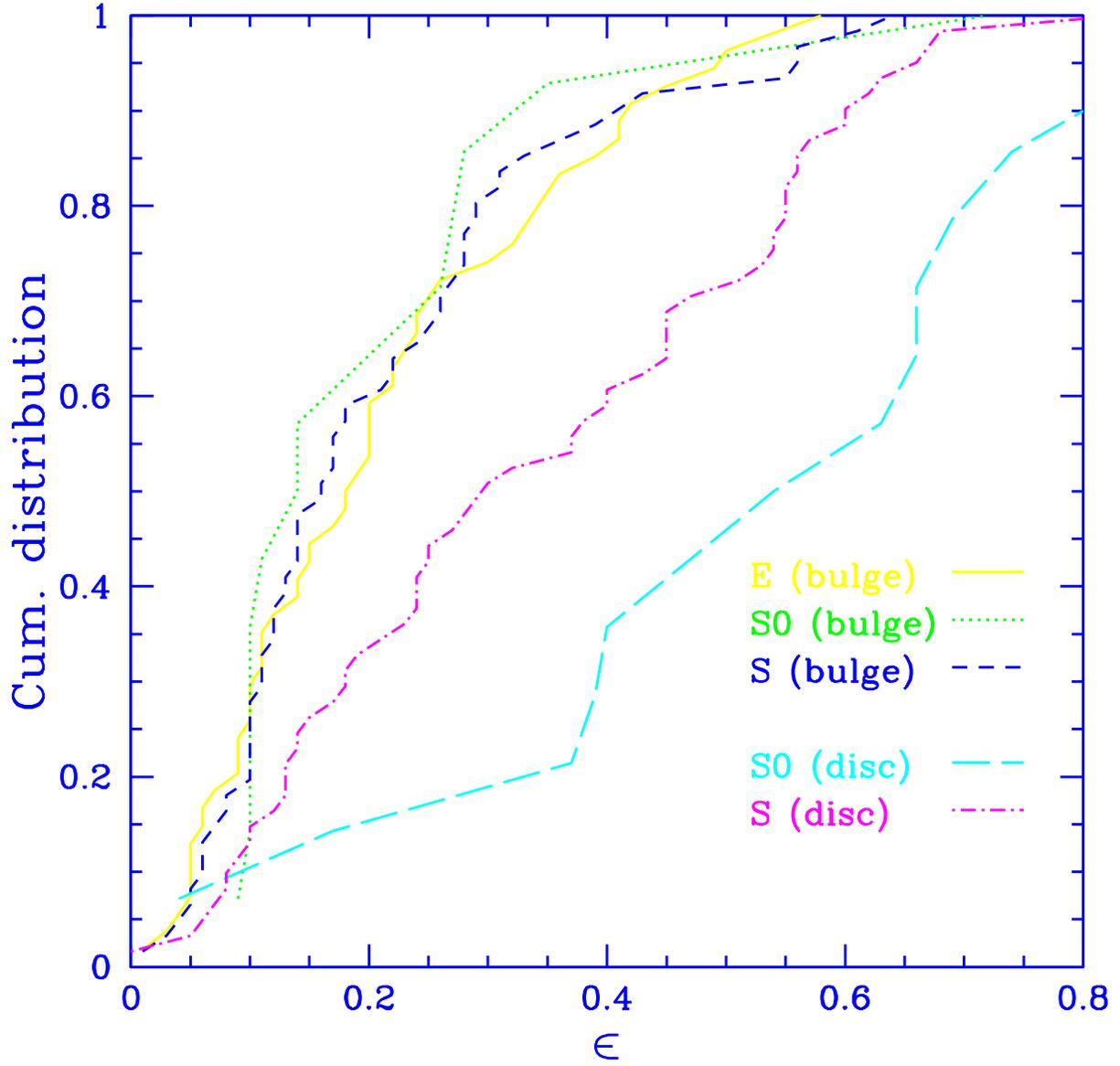}
\caption{Cumulative distribution of projected ellipticities for 
the different galaxy components and morphological types.
\label{fig_elip}}
\end{figure}

Figure~\ref{fig_edeb} shows the seeing-corrected, 
projected ellipticities for the bulges and discs of the galaxies in our 
sample that have these two component. The ellipticity for a thin disc is 
related to the inclination angle, $i$, by:  $\cos(i)=1-\epsilon$. Clearly the
ellipticity of discs tends to be higher than those of bulges.
Figure~\ref{fig_elip} shows the cumulative distributions of ellipticities for 
the different morphological types. Table \ref{elip} shows the number of
early- and late-type galaxies, together with the mean and 
standard deviation of the projected ellipticities. The main results are that 
bulges of all morphological types have ellipticities with a similar 
distribution, and that the discs of S0s seem to have a slightly different 
distribution than the late-type galaxies (a Kolmogorov--Smirnov test rejects both 
distributions being the same at the 86 \% confidence level). 
%Of course, the true distribution of disc inclinations in S0 galaxies 
%must be the same as in spiral galaxies --- to us, they should appear 
%randomly distributed over all angles.  The higher (on average) disc 
%ellipticities observed in the 14 S0 galaxies may indicate that we have 
%missed a number of faint face-on discs in this galaxy population.  
However, it may simply be a result of small number statistics.  
This needs to be investigated with a larger sample of S0 galaxies.

Jorgensen \& Franx (1994) measured the ellipticities of a volume-limited 
sample of galaxies in the Coma cluster.  However, they simply measured the global 
ellipticity of the galaxies without disentangling the disc and bulge 
components. 
They concluded that S0 galaxies tend to have larger ellipticities than pure 
ellipticals, and argued that this indicates that some of the faint face-on 
S0 galaxies were misclassified as E types. On the basis of our own results 
(Table~\ref{elip}), the differences in ellipticity found by those authors 
could indeed be due to the differences in ellipticity between the two structural components (bulges and discs).

\subsection{Line-of-sight velocity distributions}

\subsubsection{Coma elliptical galaxies}

Studying the velocity distributions of the different morphological types  in Coma
allows one to test models of the origin and evolution of these types. Previous
dynamical analyses of groups and clusters have been conducted by many teams. For
instance, Zabludoff \& Franx (1993) have analyzed six rich clusters of galaxies;
Schindler, Binggeli, \& Bohringer (1999) and Conselice, Gallagher, \& Wyse (2001)
have explored the Virgo cluster; Held \& Mould (1997)  and Drinkwater, Gregg, \&
Colless (2001) studied the Fornax cluster;  Colless \& Dunn (1996) the Coma
cluster; and Cote et al.\ (1997) studied the Centaurus A group. In general, all
these works show that  late-type galaxies and dwarf ellipticals have broader
velocity distributions   as compared with giant ellipticals. This has been
interpreted as evidence that  spiral and dE galaxies are infalling into a
virialized core dominated by giant ellipticals. 

The Coma cluster is not a simple relaxed cluster: a close examination of its 
central parts reveals the presence of two groups of galaxies (Fitchett \& 
Webster 1987; Baier, Fritze \& Tiersch 1990) dominated by the cD galaxies 
NGC~4874 and NGC~4889. The analysis of radial velocities by Colless \& 
Dunn (1996) found these two concentrations to be dynamically different 
entities. A third group, dominated by the galaxy NGC~4839, is located
40$^{\prime}$ to the SW of the cluster. Burns et al.\ (1994) claimed that the 
group had already been disrupted after its first passage through the cluster, 
while Colless \& Dunn (1996) suggested that this group is falling into Coma 
along the Great Wall. 
The Coma cluster is one the brightest extragalactic X-ray sources observed by 
{\itshape ROSAT} (White et al.\ 1993), {\itshape ASCA} (Watanabe et al.
\ 1999), and {\itshape XMM} (Briel et al.\ 2001). In the X-rays images, 
the cluster appears elongated along the line connecting NGC~4874 and 
NGC~4889, and numerous lumps and individual sources are visible. 
Based on these observations, Neumann et al.\ (2001) discussed the morphology
of the NGC~4839 group and concluded that the group is falling into Coma for 
the first time, in agreement with the scenario proposed by Colless \&
Dunn (1996).
Gurzadyan \& Mazure (2001) analysed the 
substructure of the Coma cluster using an S-tree method and also concluded 
that three subgroups exist. Furthermore, a study of the small scale structure 
conducted by Conselice \& Gallagher (1998) discovered three additional
aggregates, two of them equidistant between NGC~4874 and NGC~4889, and the 
other near the giant elliptical NGC~4860.

\begin{figure}  
\plotone{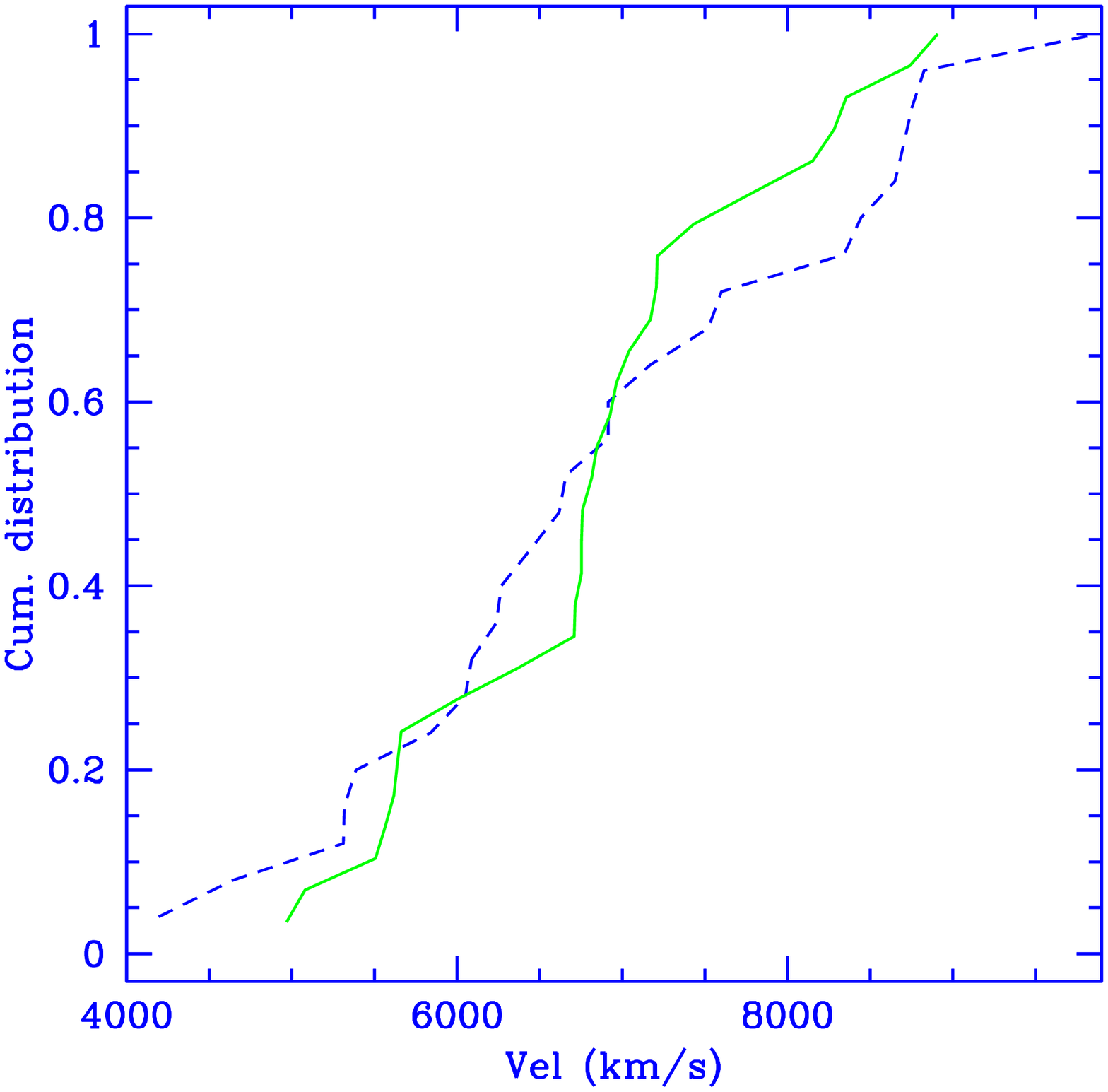}
\caption{Velocity distribution for ellipticals in Coma 
with $n\ge 2$ (\textit{solid}) and $n\le 2$ (\textit{dashed}).
\label{fig_vel}}
\end{figure}

Edwards et al.\ (2002) have conducted a detailed kinematical study of Coma. 
Analyzing differences in velocity between the giant and dwarf galaxies (based on whether 
they are brighter or fainter than $M_B = -16.84$ mag), they found that the
giants follow a non-Gaussian distribution in velocity, while the distribution 
of the dwarfs is compatible with a Gaussian. 
While we use the same compilation of redshifts, our 
quantitative morphological analysis allows us to study the motions of the 
galaxies as a function of their structural parameters. The velocity
distributions for ellipticals with $n< 2$ (dwarfs) and $n>2$ (giants) are shown in
Fig.~\ref{fig_vel}. Both distributions are symmetric with respect to the 
systemic velocity of the cluster. The distribution for galaxies with $n > 2$ 
is narrower and shows a larger number of galaxies with velocities close
to the cluster systemic velocity. This could indicate a concentration 
of galaxies with $n> 2$ in the central parts of the cluster, in agreement with 
previous findings that dwarf galaxies are less concentrated than bright 
galaxies in Coma (Quintana 1979).  Similar results have been found for the 
Fornax cluster (e.g., Caldwell 1987; Drinkwater et al.\ 2001), 
the cluster AC118 at $z\sim 0.3$ (Andreon 2002), and for a sample of other clusters 
between $z=0$ and $z=0.5$ (Adami et al.\ 2001).  Table~\ref{velo} summarizes this 
analysis showing the mean velocity (and dispersion) obtained for the different 
types of galaxies discussed above.

A Kolmogorov--Smirnov test applied to the velocity distributions of E galaxies
with $n> 2$ and  $n< 2$ shows some evidence (70\% of confidence level) for the
two distributions being different. Table~\ref{velo} shows that the ratio 
between cluster velocity dispersion for these galaxies is $\sim 4/3$.  
Edwards et al.\ also  found a slightly larger dispersion for dwarf than for 
giant galaxies  (1096$\pm$ 45 and $979\pm$ 30 km s$^{-1}$ respectively) in 
Coma.
% but this difference is smaller than the one that
%we have found between early-type galaxies with $n>2$ and $n<2$. This
%indicates that the segregation is more related to
%structural  parameters than to magnitude.

If the central 0.28 square degrees of the Coma cluster were fully relaxed, 
by dynamical relaxation (Binney \& Tremaine, 1987), this should correspond to 
mass ratios between the two groups of galaxies $\sim 16/9$. Assuming a similar
mass-to-light ratio for all the ellipticals, this would imply a mean 
difference in magnitude between both  groups  $\sim 0.6$, which is smaller 
than the actual difference (2.4 mag). For instance, Edwards et al.\ have 
performed Monte Carlo simulation of the velocities in a cluster fully relaxed,
finding that the ratio between the velocity dispersion for dwarfs and giants 
should be $\sim 3$. 
Edwards et al.\, argued that the differences in velocity between the two
groups of ellipticals are the result of the merging of subclusters which were
partially relaxed. In the particular case of the Coma cluster, this could
be supported also by the large scale distribution of the groups associated
to NGC~4874 and NGC~4889 respectively.

\subsubsection{Background galaxies}

\begin{figure}  
\plotone{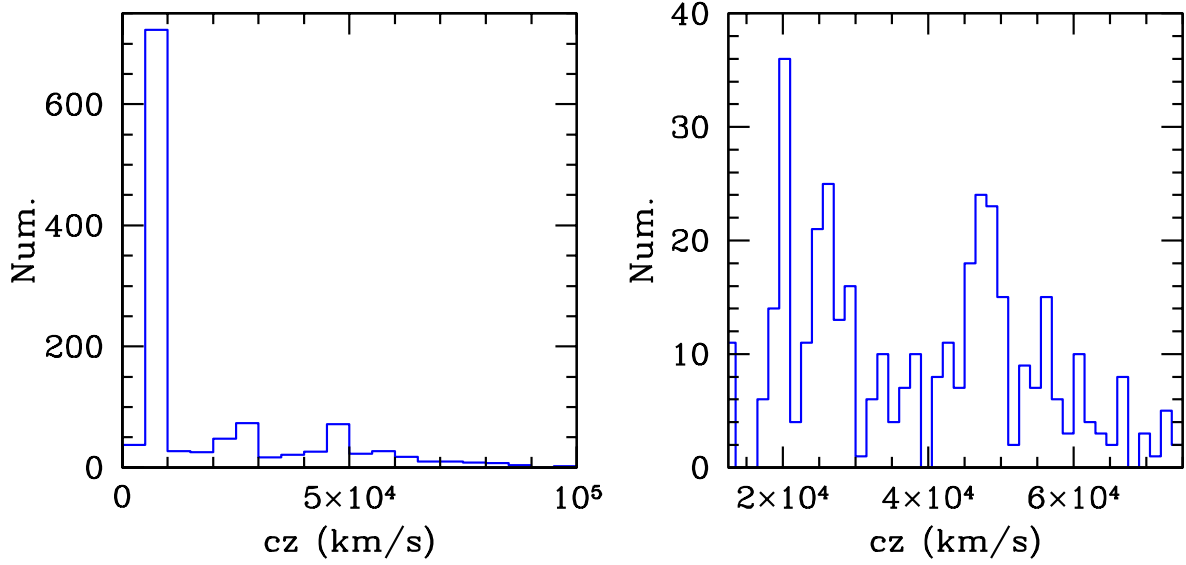}
\caption{Redshift distribution of objects in the Edwards et
al.\ (2002) compilation along the line of sight to the Coma cluster.
\textit{Left}: Full range in $z$; \textit{Right}: Only galaxies behind the 
Coma cluster.
\label{fig_histo}}
\end{figure}

In the compilation of redshifts by Edwards et al.\ (2002) there are 745 Coma
members; the rest are 421 background and 15 foreground  objects.
Figure~\ref{fig_histo} shows a histogram with the distribution of redshifts
in this catalog. The diagram on the left is dominated by a big bump at 
$\sim 7,000$ km~s$^{-1}$, which corresponds to the Coma cluster. 
There are two other concentrations at higher redshift, peaked at 
$\sim 25,000$ and $\sim 47,000$ km~s$^{-1}$; 
these can be seen in more detail in the diagram on the right of 
Figure~\ref{fig_histo}. The contrast in galaxy density between these 
structures and the average density distribution is $\sim 3$. Figure~\ref{fig_histo} also 
shows a low density region  from 10,000 km~s$^{-1}$ to 20,000 km~s$^{-1}$ that 
has been discussed in Lindner et al.\ (1995).

As the limiting magnitude of the sample with measured redshifts is unclear, 
it is not possible to make a comparison with  the values reported by Arnouts 
et al.\ (1997) for the expected density of galaxies in the field. The range of
velocities in both of the high-density structures beyond Coma are too high for 
those expected from a typical 
cluster; however, both structures show evidence of substructure within them. 
The closer concentration can be split into two substructures, one at $\sim 20,000$  
and the other at $\sim 25,000$ km~s$^{-1}$, both having velocity dispersion $\sim
1,000$ km~s$^{-1}$. Although it is difficult to appreciate in the figure,
we tentatively identify in the more distant concentration of galaxies two
overlapping structures at $\sim 46,500$ and $\sim 49,000$ km~s$^{-1}$, with
dispersions of $\sim 2,500$ and $\sim 1,000$ km~s$^{-1}$ respectively. It is 
interesting to note that both structures seem to extend over the full spatial 
region ($\sim 2$ square degrees) of the velocity catalog without a clear spatial 
concentration; they subtend angles much larger than those expected for a 
cluster at these redshifts. For instance, a typical cluster with size of 1 
Mpc would subtend an angle of $\sim 20$ arcminutes at 
$\sim 30,000$ km~s$^{-1}$. The three dimensional  structure of the background
galaxies therefore resembles large, low density regions (voids) separated by 
thin walls.

\section{Summary}

\begin{enumerate}

\item We have determined the morphological types and structural parameters of
187 galaxies in the central part of the Coma  cluster (0.28 square degrees).
The analysis extends down to $R=17$ mag. 
% for galaxies cataloged in the compilation by Edwards et al.\ (2002).  
The results of our quantitative-based ($B/T$ ratio) galaxy type determinations 
are in good agreement with previous subjective classifications. 

\item We have shown various correlations between the surface brightness, 
effective radius, luminosity, and the S\'ersic shape index $n$ for 
ellipticals and the bulges of spirals. 
% and scale length and central intensity for the discs) of the Coma cluster galaxies, 
All results are in agreement with past studies of elliptical galaxies in the
Virgo and the Fornax cluster. The strongest correlation that between the 
S\'ersic index, $n$, and luminosity. 

\item The bulges of all morphological types show a similar distribution of
projected ellipticities, while there is marginal evidence that the discs of S0 
and spirals have different projected-ellipticity distributions.

\item We have not found bright spiral galaxies ($M_R\le -22$ mag) in
the core of Coma. In the magnitude interval  $-22\le M_R\le -20$, the scale 
length of the discs of Coma spiral galaxies are 30\% smaller than for field galaxies. 
This may be evidence of environment-driven evolution. 

%\item We have analyzed the spatial structure in the background of Coma up to
%$z\sim 0.2$, showing the existence of low density regions, and thin extended
%structures with high concentration of galaxies.

\end{enumerate}

\acknowledgments

We thank A. Aparicio and A. Mar\'\i n who kindly provided us with the images 
of Coma. We also thanks to M. Colless who has provided us with the compilation of redshifts. The Isaac Newton Telescope (INT) is operated on the island
of La Palma by the Isaac Newton Group and the Instituto de Astrof\'\i sica de
Canarias (IAC) in the Spanish Observatory Roque de Los Muchachos of the IAC.

\clearpage

%\onecolumn
{\scriptsize
\begin{longtable}{llllrrrlllllrlr}

\caption{Structural parameters of galaxies analyzed in this paper. The
parameters quoted are: 
1) The identification of the galaxy according to the notation used by Godwin 
et al.\ 1983; 
2) the identification of the galaxies according with the NGC/IC and Dressler 
1980 catalogs respectively; 
3) $B$-band magnitude from Godwin et al.\ 1983; 
4) the color $B-R$ (Godwin et al.\ 1983); 
5) the recessional velocity in km s$^{-1}$ (0 if unavailable) from 
Edwards et al.\ 2002;  
6--7) the position $X-Y$ (in arcminutes from the center and using the 
usual convention $X$+ is E, $Y$+ is N); 
8) The surface brightness of the bulge at the effective radius 
(mag~arcsecs$^{-2}$ in the $R$-band); 
9) the effective  radius of the bulge (arcsec); 
10) the S\'ersic index of the bulge; 
11) the ellipticity of the bulge; 
12) the central surface brightness of the disc (mag~arcsecs$^{-2}$  in the 
R-band); 
13) the scale length of the disc (arcsec); 
14) the ellipticity of the disc; 
15) the bulge-to-total luminosity ratio ($B/T<0$ indicates inaccurate
determination of the parameters due to irregularities or the presence of 
structures such as bars, rings, etc.)} \\

GMP      & &     B     &   B-R    &v  & X  &Y &$\mu_e$  &$r_e$ &n   &$\epsilon_{bulge}$ & $\mu_0$ & h &$\epsilon_{disc}$& B/T   \\
%	 & &           &    & (km/s) &(') & (')	 &(mag~''$^{-2})$ & ('') & & & (mag~''$^{-2})$  & ('') & &    \\
\\

\hline
\\
 2347 &	    D098   &  15.85  &  1.91 &  6848 &  16.90 &  -4.27 &  18.04 &   0.70 &   1.29 &  0.11 & 18.82 &   4.22 &   0.47 &   0.17  \\
 2374 &   N4911    D082   &  13.91  &  1.67 &  7987 &  16.15 & -10.85 &  18.63 &   1.10 &   2.24 &  0.08 & 19.37 &   9.25 &   0.44 &  -0.11  \\
 2376 &		   &  18.23  &  1.77 &  6049 &  16.15 &  -4.36 &  23.16 &   6.70 &   1.83 &  0.45 & \nodata &  \nodata &  \nodata &   1.00  \\
 2385 &		   &  17.62  &  1.82 &  7040 &   5.90 &  -7.74 &  20.37 &   2.15 &   2.19 &  0.25 & \nodata &  \nodata &  \nodata &   1.00  \\
 2390 &   IC4051   D143   &  14.47  &  1.82 &  4968 &   5.82 &   2.19 &  21.79 &  15.80 &   3.86 &  0.36 & \nodata &  \nodata &  \nodata &   1.00  \\
 2393 &	    D062   &  16.51  &  1.90 &  8302 &   5.75 & -11.25 &  20.99 &   3.28 &   4.56 &  0.27 & 21.04 &   7.53 &   0.74 &   0.68  \\
 2399 &		   &  18.78  &  9.99 &  5837 &   5.68 &  -0.15 &  22.09 &   2.95 &   1.35 &  0.32 & \nodata &  \nodata &  \nodata &   1.00  \\
 2408 &		   &  18.49  &  1.73 &       &   5.46 &  -8.72 &  22.97 &   3.15 &   0.83 &  0.45 & 23.71 &   7.90 &   0.61 &   0.43  \\
 2417 &   N4908    D167   &  14.91  &  1.87 &  8742 &   5.17 &   4.34 &  20.34 &   6.96 &   3.81 &  0.34 & \nodata &  \nodata &  \nodata &   1.00  \\
 2421 &		   &  17.98  &  1.90 &  8150 &   5.10 & -13.68 &  22.01 &   1.83 &   1.17 &  0.25 & 20.96 &   4.18 &   0.53 &   0.19  \\
 2440 &   IC4045   D168   &  15.17  &  1.85 &  6899 &   4.54 &   7.18 &  21.05 &   1.83 &   1.68 &  0.30 & \nodata &  \nodata &  \nodata &  -1.00  \\
 2441 &   N4907    D205   &  14.65  &  1.74 &  5773 &   4.51 &  11.23 &  19.71 &   2.01 &   3.20 &  0.11 & 19.35 &   8.86 &   0.06 &  -0.10  \\
 2457 &	    D117   &  16.56  &  1.88 &  8604 &   4.27 &  -2.94 &  18.48 &   0.66 &   1.28 &  0.29 & 18.95 &   3.10 &   0.46 &  -0.16  \\
 2478 &		   &  18.09  &  1.86 &  8752 &   3.84 &  -8.13 &  21.56 &   3.43 &   1.35 &  0.41 & \nodata &  \nodata &  \nodata &   1.00  \\
 2489 &	    D191   &  16.69  &  1.77 &  6582 &   3.65 &   7.78 &  18.02 &   0.66 &   0.79 &  0.28 & 18.48 &   2.74 &   0.55 &   0.20  \\
 2510 &	    D116   &  16.13  &  1.90 &  8353 &   3.24 &  -0.47 &  19.11 &   1.44 &   2.31 &  0.15 & 19.92 &   4.51 &   0.30 &  -0.41  \\
 2516 &   IC4042   D144   &  15.34  &  1.86 &  6354 &   3.19 &   0.03 &  19.84 &   3.68 &   3.62 &  0.16 & 23.19 &  51.89 &   0.18 &   0.28  \\
 2519 &		   &  18.68  &  1.72 &  6216 &   3.21 &   8.72 &  22.15 &   1.24 &   1.76 &  0.21 & 22.44 &   5.55 &   0.21 &   0.14  \\
 2529 &		   &  18.63  &  1.87 &  8826 &   2.89 &   4.45 &  21.05 &   1.83 &   1.68 &  0.10 & \nodata &  \nodata &  \nodata &   1.00  \\
 2531 &		   &  17.63  &  1.70 & 18442 &   2.89 &  -4.55 &  22.77 &   0.96 &  11.43 &  0.31 & 20.14 &   3.53 &   0.23 &  -0.03  \\
 2535 &  IC4041    D145   &  15.93  &  1.90 &  7059 &   2.82 &   1.54 &  19.32 &   1.65 &   1.87 &  0.20 & 19.39 &   4.61 &   0.49 &  -0.35  \\
 2541 &   N4906    D118   &  15.44  &  1.98 &  7497 &   2.54 &  -2.82 &  19.55 &   2.59 &   2.84 &  0.20 & 19.86 &   4.58 &   0.04 &   0.52  \\
 2559 &  IC4040    D169   &  15.44  &  9.99 &  7801 &   2.08 &   5.25 &  23.00 &   3.49 &   2.41 &  0.12 & 18.68 &   4.72 &   0.83 &   0.13  \\
 2584 &	    D192   &  16.14  &  1.79 &  5465 &   1.63 &  10.52 &  20.28 &   1.10 &   5.90 &  0.05 & 18.63 &   4.91 &   0.68 &   0.12  \\
 2585 &		   &  18.44  &  1.73 &  6914 &   1.62 &  -1.69 &  22.31 &   4.12 &   1.82 &  0.41 & \nodata &  \nodata &  \nodata &   1.00  \\
 2591 &	    D051   &  18.50  &  1.86 &  8356 &   1.41 &  -2.17 &  22.94 &   4.11 &   2.13 &  0.09 & \nodata &  \nodata &  \nodata &   1.00  \\
 2603 &		   &  17.36  &  1.80 &  8152 &   1.17 &  -8.81 &  21.60 &   4.74 &   2.15 &  0.50 & \nodata &  \nodata &  \nodata &   1.00  \\
 2615 &	    D063   &  16.97  &  1.90 &  6679 &   0.99 & -12.29 &  19.30 &   0.90 &   1.02 &  0.24 & 19.74 &   3.46 &   0.33 &  -0.18  \\
 2619 &		   &  18.38  &  1.71 &  6279 &   0.97 &  17.10 &  22.16 &   1.19 &   1.04 &  0.27 & 21.21 &   3.85 &   0.24 &   0.07  \\
 2633 &		   &  18.48  &  1.75 &  -359 &   0.48 &  11.37 &  20.78 &   0.63 &   2.38 &  0.37 & 20.37 &   2.30 &   0.34 &   0.12  \\
 2651 &	    D147   &  16.19  &  1.85 &  7700 &   0.07 &   0.08 &  19.44 &   1.42 &   2.33 &  0.39 & 19.82 &   5.53 &   0.54 &  -0.26  \\
 2654 &	    D119   &  16.38  &  1.90 &  6984 &   9.98 &  -0.90 &  17.90 &   0.73 &   1.79 &  0.08 & 19.86 &   4.12 &   0.37 &   0.41  \\
 2692 &		   &  18.20  &  1.78 &  7962 &   9.28 &  -2.66 &  23.76 &   4.94 &   3.55 &  0.14 & 20.89 &   3.66 &   0.66 &   0.52  \\
 2711 &		   &  18.90  &  9.99 & 19792 &   8.90 &  14.20 &  19.49 &   0.60 &   1.96 &  0.21 & \nodata &  \nodata &  \nodata &   1.00  \\
 2727 &   IC4026   D170   &  15.73  &  1.77 &  8161 &   8.69 &   4.56 &  18.52 &   1.10 &   1.76 &  0.15 & 19.43 &   4.20 &   0.20 &  -0.29  \\
 2728 &		   &  18.15  &  1.73 &  7965 &   8.70 &  16.57 &  20.21 &   0.38 &   1.52 &  0.10 & 19.53 &   2.69 &   0.56 &   0.05  \\
 2736 &		   &  18.21  &  1.76 &  4869 &   8.59 &  -4.34 &  19.84 &   0.63 &   0.62 &  0.28 & 20.37 &   2.19 &   0.25 &   0.17  \\
 2753 &		   &  18.10  &  1.86 &  7767 &   8.26 &  -7.65 &  21.40 &   2.33 &   0.90 &  0.22 & 22.06 &   4.45 &   0.23 &   0.48  \\
 2777 &		   &  18.21  &  1.71 &  6202 &   7.97 &   2.30 &  19.17 &   0.74 &   2.64 &  0.06 & \nodata &  \nodata &  \nodata &  -1.00  \\
 2778 &		   &  16.69  &  1.81 &  5122 &   7.95 &  -2.03 &  20.43 &   1.08 &   1.25 &  0.38 & 19.27 &   2.86 &   0.05 &  -0.06  \\
 2783 &		   &  17.37  &  1.83 &  5318 &   7.90 &  -9.33 &  23.13 &   8.44 &   2.64 &  0.08 & \nodata &  \nodata &  \nodata &  -1.00  \\
 2784 &		   &  18.36  &  1.81 &  7838 &   7.90 &   7.57 &  21.38 &   1.64 &   0.97 &  0.13 & 22.05 &   4.54 &   0.19 &   0.33  \\
 2787 &		   &  18.46  &  1.65 &  9873 &   7.85 &   5.30 &  22.46 &   3.79 &   1.71 &  0.15 & \nodata &  \nodata &  \nodata &   1.00  \\
 2795 &   N4895    D206   &  14.38  &  9.99 &  8469 &   7.76 &  13.89 &  20.19 &   8.84 &   2.49 &  0.10 & 21.20 &  23.80 &   0.69 &   0.74  \\
 2799 &		   &  18.70  &  9.99 &  6136 &   7.70 &   1.00 &  21.27 &   1.39 &   1.73 &  0.07 & \nodata &  \nodata &  \nodata &  -1.00  \\
 2800 &		   &  18.38  &  9.99 &  7015 &   7.70 & -11.19 &  22.77 &   1.78 &   1.80 &  0.05 & 22.37 &   5.19 &   0.13 &   0.18  \\
 2815 &   N4894    D122   &  15.87  &  1.74 &  4636 &   7.45 &  -0.20 &  17.96 &   0.73 &   1.52 &  0.12 & 19.15 &   4.25 &   0.60 &   0.31  \\
 2839 &   IC4021   D172   &  16.01  &  1.75 &  5724 &   7.06 &   4.22 &  17.37 &   0.69 &   1.56 &  0.22 & 18.94 &   2.73 &   0.10 &   0.35  \\
 2852 &		   &  17.80  &  1.79 &  7405 &   6.81 &  -6.22 &  21.00 &   0.88 &   1.49 &  0.29 & 20.78 &   3.04 &   0.24 &   0.13  \\
 2856 &		   &  18.23  &  1.57 &  8134 &   6.77 &   4.94 &  23.83 &  11.98 &   1.29 &  0.72 & 18.96 &   1.63 &   0.92 &   0.82  \\
 2861 &	    D173   &  16.26  &  1.85 &  7511 &   6.65 &   6.27 &  18.88 &   1.42 &   2.20 &  0.12 & 20.21 &   4.81 &   0.28 &   0.49  \\
 2866 &	    D064   &  16.90  &  1.79 &  7030 &   6.59 & -11.34 &  20.69 &   1.84 &   2.61 &  0.11 & 21.49 &   4.71 &   0.05 &   0.47  \\
 2895 &	           &  18.64  &  2.04 & 26236 &   6.05 &  -0.77 &  21.78 &   0.69 &   0.83 &  0.28 & 20.17 &   2.10 &   0.56 &   0.06  \\
 2897 &	    D099   &  16.98  &  1.53 &  9889 &   6.00 &  -6.42 &  21.67 &   2.03 &   5.37 &  0.64 & 19.89 &   2.48 &   0.08 &   0.17  \\
 2910 &	    D100   &  16.25  &  1.41 &  5136 &   5.78 &  -6.28 &  20.58 &   3.54 &   1.82 &  0.30 & \nodata &  \nodata &  \nodata &  -1.00  \\
 2912 &   N4895A   D207   &  16.07  &  1.80 &  6752 &   5.82 &  11.97 &  20.69 &   5.13 &   3.58 &  0.35 & \nodata &  \nodata &  \nodata &   1.00  \\
 2914 &		   &  17.18  &  1.81 &  7436 &   5.74 &  11.34 &  20.27 &   0.97 &   1.35 &  0.06 & 20.55 &   4.03 &   0.14 &   0.15  \\
 2921 &    N4889   D148   &  12.62  &  1.91 &  6508 &   5.57 &   0.35 &  19.17 &   2.62 &   1.88 &  0.34 & \nodata &  \nodata &  \nodata &   1.00  \\
 2922 &   IC4012   D174   &  15.93  &  1.86 &  7207 &   5.57 &   6.46 &  19.60 &   3.08 &   3.26 &  0.20 & \nodata &  \nodata &  \nodata &   1.00  \\
 2923 &		   &  17.65  &  9.99 &  8664 &   5.57 & -11.86 &  21.40 &   2.87 &   0.97 &  0.43 & 22.17 &   7.12 &   0.62 &   0.48  \\
 2929 &		   &  18.66  &  1.90 &  6266 &   5.45 &  -0.77 &  22.94 &   3.49 &   1.73 &  0.05 & \nodata &  \nodata &  \nodata &   1.00  \\
 2940 &   IC4011   D150   &  16.08  &  1.82 &  7244 &   5.21 &   2.00 &  17.21 &   0.42 &   1.32 &  0.06 & 18.78 &   2.31 &   0.09 &   0.24  \\
 2943 &		   &  17.66  &  9.99 & 13298 &   5.18 &  16.84 &  24.98 &   0.81 &  67.60 &  0.17 & 21.45 &   5.48 &   0.12 &   0.01  \\
 2945 &	    D065   &  16.15  &  1.77 &  6150 &   5.14 & -11.73 &  19.27 &   1.20 &   1.46 &  0.17 & 19.74 &   4.73 &   0.55 &   0.29  \\
 2956 &	    D084   &  16.20  &  1.98 &  6566 &   5.02 &  -9.79 &  22.74 &  12.49 &   4.67 &  0.10 & 19.63 &   3.61 &   0.63 &   0.86  \\
 2960 &		   &  16.78  &  1.74 &  5827 &   5.00 &   3.22 &  20.03 &   0.99 &   2.05 &  0.14 & 19.73 &   3.78 &   0.63 &   0.24  \\
 2975 &   N4886    D151   &  14.83  &  1.76 &  6363 &   4.78 &   1.01 &  21.39 &  16.37 &   7.00 &  0.05 & \nodata &  \nodata &  \nodata &   1.00  \\
 2976 &		   &  18.14  &  1.94 &  6693 &   4.75 &  11.09 &  23.06 &   6.04 &   1.36 &  0.45 & \nodata &  \nodata &  \nodata &  -1.00  \\
 2985 &		   &  17.87  &  1.65 &  5317 &   4.63 &  -0.37 &  21.96 &   3.25 &   1.65 &  0.33 & \nodata &  \nodata &  \nodata &   1.00  \\
 2989 &		   &  17.05  &  9.99 &  7675 &   4.48 &  16.17 &  18.42 &   0.57 &   1.70 &  0.18 & 20.75 &   3.81 &   0.29 &   0.35  \\
 3012 &		   &  17.49  &  1.83 &  8033 &   4.13 & -14.41 &  22.29 &   2.20 &   1.65 &  0.14 & 21.18 &   5.41 &   0.51 &   0.20  \\
 3017 &		   &  17.91  &  1.65 &  6965 &   4.01 &  -1.52 &  23.05 &   4.53 &   3.92 &  0.15 & \nodata &  \nodata &  \nodata &   1.00  \\
 3034 &		   &  18.06  &  1.70 &  5994 &   3.69 &  -1.82 &  23.72 &   6.90 &   2.72 &  0.10 & \nodata &  \nodata &  \nodata &   1.00  \\
 3055 &   N4881    D217   &  14.73  &  1.87 &  6757 &   3.32 &  16.56 &  21.74 &  12.55 &   4.44 &  0.05 & \nodata &  \nodata &  \nodata &   1.00  \\
 3068 &	    D123   &  16.47  &  1.93 &  7716 &   3.07 &  -2.43 &  19.22 &   0.80 &   1.83 &  0.29 & 19.44 &   3.89 &   0.45 &  -0.14  \\
 3071 &		   &  17.17  &  1.18 &  8909 &   2.95 & -13.46 &  21.56 &   4.40 &   2.22 &  0.10 & 21.60 &   0.55 &   0.69 &   1.00  \\
 3073 &   N4883    D175   &  15.43  &  1.89 &  8130 &   2.93 &   3.83 &  18.45 &   1.24 &   1.45 &  0.13 & 18.98 &   4.16 &   0.25 &   0.27  \\
 3084 &	    D193   &  16.43  &  1.82 &  7538 &   2.73 &   9.45 &  17.59 &   0.51 &   0.75 &  0.13 & 18.76 &   2.25 &   0.15 &  -0.21  \\
 3092 &	    D085   &  17.55  &  1.59 &  8094 &   2.66 & -10.49 &  19.95 &   0.79 &   2.15 &  0.04 & 19.90 &   1.87 &   0.12 &   0.33  \\
 3098 &		   &  18.63  &  1.74 &  6618 &   2.46 &  -0.02 &  22.57 &   3.06 &   1.62 &  0.09 & \nodata &  \nodata &  \nodata &   1.00  \\
 3113 &		   &  17.82  &  1.81 &  7601 &   1.99 &   7.66 &  21.54 &   3.26 &   1.88 &  0.24 & \nodata &  \nodata &  \nodata &   1.00  \\
 3121 &		   &  17.34  &  1.76 &  7418 &   1.93 &   6.16 &  20.38 &   0.55 &   0.92 &  0.07 & 20.36 &   3.12 &   0.10 &   0.06  \\
 3126 &		   &  17.55  &  1.82 &  7906 &   1.81 &  -8.27 &  20.17 &   0.52 &   3.12 &  0.61 & 19.46 &   2.73 &   0.60 &   0.06  \\
 3129 &		   &  17.94  &  1.71 &  6815 &   1.67 &  10.41 &  21.81 &   4.48 &   2.22 &  0.54 & \nodata &  \nodata &  \nodata &   1.00  \\
 3133 &		   &  17.23  &  1.92 &  9770 &   1.62 &  -2.76 &  19.30 &   1.01 &   0.80 &  0.24 & 19.87 &   2.57 &   0.27 &   0.32  \\
 3146 &		   &  18.54  &  1.58 &  5312 &   1.29 &   0.72 &  22.80 &   3.22 &   1.12 &  0.09 & \nodata &  \nodata &  \nodata &   1.00  \\
 3154 &		   &  18.64  &  1.68 &       &   1.12 &   7.63 &  21.60 &   2.11 &   1.31 &  0.19 & \nodata &  \nodata &  \nodata &   1.00  \\
 3160 &		   &  18.88  &  9.99 &  9546 &   0.98 &   5.03 &  23.32 &   1.15 &   1.26 &  0.03 & 22.34 &   4.69 &   0.56 &   0.10  \\
 3166 &		   &  18.37  &  1.79 &  8443 &   0.92 &   1.27 &  22.73 &   3.75 &   1.98 &  0.22 & \nodata &  \nodata &  \nodata &   1.00  \\
 3170 &   IC3998   D152   &  15.70  &  1.90 &  9399 &   0.88 &   0.19 &  19.33 &   1.62 &   3.09 &  0.27 & 19.71 &   4.89 &   0.25 &  -0.32  \\
 3172 &	           &  18.69  &  1.83 & 27401 &   0.86 &  18.25 &  21.35 &   1.67 &   4.86 &  0.11 & \nodata &  \nodata &  \nodata &   1.00  \\
 3178 &	    D101   &  16.18  &  1.84 &  7991 &   0.73 &  -6.81 &  20.97 &   3.76 &   4.69 &  0.10 & 20.38 &   4.25 &   0.66 &   0.82  \\
 3192 &		   &  18.17  &  1.49 &  5959 &   0.48 & -14.02 &  21.67 &   0.83 &   1.28 &  0.01 & 21.80 &   4.11 &   0.15 &   0.10  \\
 3196 &		   &  18.35  &  1.82 &  6847 &   0.41 &  -4.86 &  21.91 &   2.40 &   2.02 &  0.18 & \nodata &  \nodata &  \nodata &   1.00  \\
 3204 &		   &  18.54  &  1.88 &  8340 &   0.32 &  12.33 &  21.02 &   1.71 &   1.87 &  0.11 & \nodata &  \nodata &  \nodata &   1.00  \\
 3205 &		   &  17.61  &  1.83 &  6237 &   0.31 &  -6.19 &  21.56 &   3.59 &   1.62 &  0.30 & \nodata &  \nodata &  \nodata &   1.00  \\
 3206 &	    D126   &  16.36  &  1.79 &  6869 &   0.31 &  -0.74 &  19.88 &   0.90 &   3.64 &  0.14 & 18.80 &   2.67 &   0.45 &   0.18  \\
 3213 &	    D153   &  16.14  &  1.83 &  6752 &   0.21 &   1.43 &  20.57 &   3.32 &   4.14 &  0.06 & \nodata &  \nodata &  \nodata &   1.00  \\
 3222 &	    D125   &  16.47  &  1.75 &  6925 &  -0.11 &  -2.76 &  19.53 &   1.89 &   4.77 &  0.12 & \nodata &  \nodata &  \nodata &   1.00  \\
 3225 &		   &  18.33  &  1.70 & 16911 &  -0.13 &  -6.71 &  19.97 &   0.82 &   0.95 &  0.20 & 20.31 &   2.59 &   0.60 &   0.34  \\
 3245 &		   &  18.42  &  2.18 & 29670 &  -0.42 &  12.87 &  20.55 &   1.68 &   3.20 &  0.33 & 24.44 &  15.61 &   0.20 &   0.53  \\
 3248 &		   &  18.75  &  1.78 &  7519 &  -0.43 &  10.43 &  22.38 &   3.09 &   1.85 &  0.11 & \nodata &  \nodata &  \nodata &   1.00  \\
 3254 &	    D127   &  16.57  &  1.84 &  7549 &  -0.56 &  -0.15 &  19.56 &   1.09 &   3.41 &  0.26 & 19.97 &   3.14 &   0.45 &   0.44  \\
 3258 &		   &  18.61  &  1.68 & 32977 &  -0.56 &  12.52 &  18.01 &   0.36 &   1.87 &  0.83 & 19.51 &   1.37 &   0.34 &   0.15  \\
 3262 &	    D102   &  16.77  &  1.82 &  3690 &  -0.64 &  -6.96 &  21.79 &   5.25 &   3.55 &  0.32 & \nodata &  \nodata &  \nodata &  -1.00  \\
 3269 &	    D128   &  16.12  &  1.75 &  7860 &  -0.69 &  -1.01 &  18.07 &   0.60 &   1.79 &  0.28 & 18.95 &   2.77 &   0.57 &   0.31  \\
 3291 &	    D154   &  16.41  &  1.78 &  6927 &  -0.98 &   0.99 &  22.03 &   4.19 &   3.62 &  0.26 & 20.80 &   4.37 &   0.39 &   0.55  \\
 3292 &		   &  17.70  &  1.85 &  4890 &  -1.05 &   1.82 &  19.90 &   0.47 &   1.17 &  0.06 & 19.94 &   2.02 &   0.18 &   0.12  \\
 3296 &   N4875    D104   &  15.88  &  1.96 &  8005 &  -1.08 &  -3.80 &  18.55 &   1.12 &   2.32 &  0.13 & 19.34 &   3.21 &   0.21 &  -0.43  \\
 3298 &		   &  17.26  &  1.79 &  6833 &  -1.11 & -11.63 &  21.30 &   1.60 &   1.48 &  0.41 & 21.19 &   6.86 &   0.55 &   0.13  \\
 3302 &		   &  17.33  &  1.68 &  5617 &  -1.13 &  11.71 &  20.85 &   2.56 &   2.11 &  0.24 & \nodata &  \nodata &  \nodata &   1.00  \\
 3312 &		   &  18.68  &  1.78 &  7293 &  -1.27 &   2.88 &  20.92 &   1.29 &   1.73 &  0.09 & \nodata &  \nodata &  \nodata &  -1.00  \\
 3313 &		   &  17.53  &  1.83 &  6262 &  -1.29 &  -8.69 &  19.70 &   1.30 &   2.07 &  0.07 & \nodata &  \nodata &  \nodata &  -1.00  \\
 3329 &   N4874    D129   &  12.78  &  9.99 &  7166 &  -1.56 &  -0.68 &  19.86 &   8.02 &   1.25 &  0.15 & \nodata &  \nodata &  \nodata &   1.00  \\
 3336 &		   &  18.47  &  9.99 & 13278 &  -1.57 &  -3.88 &  21.49 &   1.60 &   2.15 &  0.14 & \nodata &  \nodata &  \nodata &   1.00  \\
 3339 &		   &  17.54  &  1.78 &  6279 &  -1.66 &  -6.43 &  20.40 &   1.69 &   1.07 &  0.11 & 22.49 &   5.14 &   0.17 &   0.61  \\
 3340 &		   &  18.54  &  1.93 &  4193 &  -1.68 &  -2.16 &  22.22 &   2.46 &   1.67 &  0.19 & \nodata &  \nodata &  \nodata &   1.00  \\
 3367 &   N4873    D155   &  15.15  &  1.91 &  5834 &  -2.20 &   0.78 &  17.85 &   0.77 &   1.67 &  0.14 & 18.80 &   3.69 &   0.24 &   0.22  \\
 3383 &		   &  18.50  &  1.86 &  4640 &  -2.41 &  -6.56 &  21.31 &   1.80 &   1.61 &  0.11 & \nodata &  \nodata &  \nodata &   1.00  \\
 3387 &		   &  18.22  &  1.70 &  7433 &  -2.45 &   7.79 &  23.23 &   4.97 &   2.30 &  0.14 & \nodata &  \nodata &  \nodata &   1.00  \\
 3390 &	    D176   &  15.89  &  1.75 &  6892 &  -2.49 &   4.56 &  17.83 &   0.79 &   1.09 &  0.31 & 18.21 &   2.74 &   0.54 &   0.26  \\
 3392 &		   &  18.07  &  1.44 &       &  -2.62 &  -0.97 &  23.09 &   6.78 &   1.49 &  0.26 & \nodata &  \nodata &  \nodata &   1.00  \\
 3400 &   IC3973   D103   &  15.32  &  1.88 &  4722 &  -2.65 &  -5.19 &  17.80 &   1.02 &   1.92 &  0.10 & 19.34 &   3.79 &   0.41 &  -0.54  \\
 3403 &	    D087   &  16.87  &  1.79 &  7790 &  -2.70 & -10.75 &  22.10 &   4.95 &   5.68 &  0.07 & \nodata &  \nodata &  \nodata &   1.00  \\
 3406 &		   &  18.76  &  1.76 &  7166 &  -2.76 &   3.01 &  21.75 &   1.93 &   1.57 &  0.17 & \nodata &  \nodata &  \nodata &   1.00  \\
 3414 &   N4871    D131   &  14.89  &  1.90 &  6788 &  -2.82 &  -0.86 &  18.60 &   1.26 &   2.59 &  0.10 & 19.43 &   5.75 &   0.40 &   0.32  \\
 3423 &   IC3976   D088   &  15.80  &  1.95 &  6819 &  -2.97 &  -7.23 &  19.40 &   1.93 &   4.27 &  0.09 & 19.25 &   3.73 &   0.40 &   0.57  \\
 3433 &	    D177   &  16.56  &  1.79 &  5542 &  -3.09 &   4.19 &  18.68 &   0.84 &   1.40 &  0.31 & 19.07 &   2.52 &   0.43 &   0.30  \\
 3439 &	    D178   &  16.72  &  1.81 &  5641 &  -3.15 &   6.89 &  20.81 &   1.37 &   4.81 &  0.10 & 19.97 &   2.65 &   0.14 &   0.33  \\
 3448 &		   &  18.39  &  2.28 & 29731 &  -3.27 &  15.57 &  20.16 &   1.02 &   5.75 &  0.20 & \nodata &  \nodata &  \nodata &  -1.00  \\
 3463 &		   &  18.09  &  1.87 &  6552 &  -3.44 & -11.13 &  21.80 &   0.95 &   2.01 &  0.56 & 19.79 &   2.92 &   0.67 &   0.05  \\
 3471 &	    D156   &  16.45  &  9.99 &  6708 &  -3.57 &   1.67 &  21.40 &   4.92 &   3.40 &  0.22 & \nodata &  \nodata &  \nodata &   1.00  \\
 3475 &		   &  17.21  &  1.70 &  7977 &  -3.63 &  19.01 &  17.89 &   0.38 &   2.12 &  0.10 & 18.96 &   2.14 &   0.38 &   0.25  \\
 3481 &		   &  17.69  &  1.68 &  7718 &  -3.72 &  12.98 &  18.92 &   0.63 &   1.69 &  0.10 & 20.55 &   2.33 &   0.06 &   0.42  \\
 3482 &		   &  18.66  &  1.85 &       &  -3.74 &   6.65 &  26.09 &  39.45 &   3.94 &  0.06 & \nodata &  \nodata &  \nodata &   1.00  \\
 3484 &	    D157   &  16.26  &  1.81 &  6119 &  -3.80 &   0.17 &  21.82 &   6.45 &   5.12 &  0.10 & 21.40 &   6.66 &   0.47 &   0.81  \\
 3486 &	           &  17.73  &  1.82 &  7840 &  -3.86 &  -2.17 &  19.71 &   0.87 &   1.36 &  0.17 & 21.23 &   4.52 &   0.08 &   0.23  \\
 3487 &	    D132   &  16.63  &  1.88 &  7675 &  -3.85 &  -0.16 &  19.62 &   1.47 &   2.47 &  0.10 & 20.68 &   7.08 &   0.45 &   0.35  \\
 3489 &		   &  17.93  &  9.99 &  5506 &  -3.91 &   1.57 &  24.57 &  15.69 &   4.22 &  0.20 & \nodata &  \nodata &  \nodata &   1.00  \\
 3493 &	    D067   &  16.50  &  1.94 &  6033 &  -3.97 & -13.91 &  18.75 &   1.30 &   1.61 &  0.28 & 19.77 &   3.54 &   0.54 &   0.56  \\
 3565 &		   &  18.44  &  1.84 &  7171 &  -5.09 &   0.18 &  22.85 &   5.13 &   2.33 &  0.39 & \nodata &  \nodata &  \nodata &   1.00  \\
 3567 &		   &  18.71  &  1.77 &       &  -5.10 &  -0.15 &  22.41 &   3.48 &   1.74 &  0.37 & \nodata &  \nodata &  \nodata &   1.00  \\
 3615 &		   &  18.82  &  2.22 &  6317 &  -5.86 &   0.95 &  20.78 &   0.70 &   3.64 &  0.26 & 20.21 &   1.70 &   0.25 &  -0.26  \\
 3639 &   N4867    D133   &  15.44  &  1.83 &  4805 &  -6.12 &   0.02 &  21.52 &   9.11 &   6.13 &  0.28 & \nodata &  \nodata &  \nodata &  -1.00  \\
 3645 &		   &  18.64  &  1.86 &  6417 &  -6.23 &  -4.50 &  20.50 &   1.25 &   1.64 &  0.08 & \nodata &  \nodata &  \nodata &  -1.00  \\
 3656 &	    D180   &  15.53  &  1.77 &  7815 &  -6.33 &   6.36 &  19.31 &   1.09 &   2.52 &  0.11 & 20.26 &   5.60 &   0.07 &   0.20  \\
 3660 &   IC3963   D068   &  15.76  &  1.87 &  6853 &  -6.49 & -11.76 &  19.45 &   1.55 &   2.65 &  0.11 & 19.96 &   5.64 &   0.35 &  -0.33  \\
 3664 &   N4864    D159   &  14.70  &  9.99 &  6812 &  -6.61 &   0.38 &  19.93 &   4.97 &   2.36 &  0.10 & \nodata &  \nodata &  \nodata &  -1.00  \\
 3681 &		   &  18.01  &  1.73 &  6913 &  -6.89 &   2.32 &  21.25 &   2.25 &   1.55 &  0.05 & \nodata &  \nodata &  \nodata &   1.00  \\
 3706 &		   &  17.61  &  1.85 &  6891 &  -7.33 &  -6.19 &  19.53 &   1.16 &   3.38 &  0.13 & \nodata &  \nodata &  \nodata &  -1.00  \\
 3707 &		   &  17.76  &  1.82 &  7212 &  -7.34 &   4.22 &  21.80 &   5.03 &   2.92 &  0.58 & \nodata &  \nodata &  \nodata &   1.00  \\
 3719 &		   &  18.29  &  1.91 &  7812 &  -7.45 &  -4.38 &  21.27 &   1.11 &   2.83 &  0.17 & 21.11 &   2.51 &   0.17 &   0.34  \\
 3730 &   IC3959   D069   &  15.27  &  1.94 &  7044 &  -7.66 & -11.18 &  19.48 &   3.24 &   2.29 &  0.13 & \nodata &  \nodata &  \nodata &  -1.00  \\
 3733 &   IC3960   D109   &  15.85  &  1.89 &  6600 &  -7.70 &  -6.93 &  17.65 &   0.75 &   1.16 &  0.18 & 18.89 &   2.97 &   0.13 &   0.28  \\
 3739 &   IC3957   D070   &  15.88  &  1.87 &  6352 &  -7.85 & -12.11 &  22.55 &   9.58 &   9.59 &  0.09 & \nodata &  \nodata &  \nodata &  -1.00  \\
 3750 &		   &  18.08  &  1.91 &  6267 &  -7.99 & -11.88 &  20.04 &   1.27 &   1.24 &  0.39 & 21.82 &   4.72 &   0.00 &   0.32  \\
 3761 &   IC3955   D160   &  15.57  &  1.88 &  7664 &  -8.11 &   1.57 &  18.43 &   0.99 &   1.73 &  0.05 & 19.43 &   4.94 &   0.29 &  -0.24  \\
 3780 &		   &  17.89  &  1.85 &  5080 &  -8.37 &   4.79 &  22.40 &   5.40 &   2.62 &  0.49 & \nodata &  \nodata &  \nodata &   1.00  \\
 3782 &	    D108   &  16.55  &  1.85 &  6426 &  -8.43 &  -3.57 &  19.45 &   1.22 &   3.31 &  0.16 & 19.50 &   3.53 &   0.32 &   0.34  \\
 3794 &	    D134   &  17.37  &  1.98 &  6956 &  -8.53 &  -0.68 &  18.82 &   0.82 &   2.05 &  0.26 & 19.21 &   1.69 &   0.13 &   0.43  \\
 3851 &	    D135   &  16.98  &  1.86 &  8283 &  -9.43 &  -0.18 &  22.18 &   6.06 &   4.92 &  0.23 & \nodata &  \nodata &  \nodata &   1.00  \\
 3855 &		   &  18.05  &  1.79 &  5564 &  -9.57 &  -2.16 &  22.54 &   4.40 &   2.25 &  0.20 & \nodata &  \nodata &  \nodata &   1.00  \\
 3882 &	    D071   &  16.97  &  1.85 &  6893 & -10.00 & -11.10 &  20.70 &   2.71 &   1.72 &  0.56 & 20.34 &   5.04 &   0.66 &   0.39  \\
 3895 &		   &  17.74  &  1.72 &  8651 & -10.21 &  -8.91 &  21.37 &   3.43 &   1.01 &  0.26 & \nodata &  \nodata &  \nodata &   1.00  \\
 3914 &	    D136   &  16.57  &  1.81 &  5661 & -10.49 &  -0.35 &  19.82 &   2.33 &   3.54 &  0.14 & \nodata &  \nodata &  \nodata &   1.00  \\
 3925 &		   &  18.00  &  2.00 &  6448 & -10.72 & -10.48 &  21.39 &   3.28 &   1.66 &  0.04 & \nodata &  \nodata &  \nodata &   1.00  \\
 3943 &		   &  16.93  &  1.88 &  5507 & -11.01 &  -9.42 &  19.86 &   1.28 &   1.33 &  0.29 & 19.69 &   3.63 &   0.37 &   0.21  \\
 3958 &   IC3947   D072   &  15.94  &  1.91 &  5655 & -11.22 & -11.12 &  19.64 &   2.31 &   3.24 &  0.35 & 19.08 &   2.86 &   0.37 &   0.57  \\
 3972 &	    D181   &  16.52  &  1.87 &  6046 & -11.46 &   6.81 &  19.40 &   1.70 &   2.32 &  0.14 & 20.51 &   5.80 &   0.84 &   0.78  \\
 3973 &		   &  18.66  &  1.86 &  6715 & -11.49 &   1.50 &  23.06 &   3.97 &   3.21 &  0.18 & \nodata &  \nodata &  \nodata &   1.00  \\
 3982 &		   &  19.40  &  2.46 & 45675 & -11.62 &   5.01 &  20.56 &   1.15 &   3.39 &  0.18 & \nodata &  \nodata &  \nodata &   1.00  \\
 3993 &		   &  19.02  &  2.36 & 47234 & -11.80 &  -3.58 &  22.18 &   2.49 &   5.72 &  0.28 & \nodata &  \nodata &  \nodata &   1.00  \\
 3997 &   IC3946   D091   &  15.28  &  1.95 &  5879 & -11.96 &  -9.60 &  17.81 &   1.30 &   1.32 &  0.36 & 19.02 &   4.96 &   0.30 &   0.29  \\
 4003 &		   &  18.60  &  1.81 &  7074 & -12.05 &   2.89 &  20.32 &   0.49 &   0.52 &  0.08 & 21.13 &   3.18 &   0.40 &   0.10  \\
 4024 &		   &  17.97  &  2.39 & 47492 & -12.43 &  -4.12 &  22.13 &   1.03 &   8.65 &  0.19 & 19.62 &   2.32 &   0.22 &   0.09  \\
 4035 &		   &  18.49  &  1.82 &  6657 & -12.67 & -12.99 &  22.94 &   4.38 &   1.97 &  0.06 & \nodata &  \nodata &  \nodata &   1.00  \\
 4060 &		   &  17.57  &  1.31 &  8703 & -13.32 & -12.59 &  22.82 &   5.69 &   1.10 &  0.20 & \nodata &  \nodata &  \nodata &   1.00  \\
 4063 &		   &  19.32  &  2.38 & 47692 & -13.45 &  -2.89 &  21.19 &   1.38 &   4.07 &  0.10 & \nodata &  \nodata &  \nodata &   1.00  \\
 4081 &		   &  18.91  &  2.15 & 50090 & -13.71 &  -3.00 &  24.95 &   9.46 &  12.19 &  0.70 & 20.48 &   1.21 &   0.32 &   0.73  \\
 4083 &		   &  17.82  &  1.91 &  6178 & -13.70 &  -8.60 &  21.12 &   2.49 &   1.65 &  0.55 & 21.14 &   4.34 &   0.54 &   0.44  \\
 4103 &		   &  17.74  &  1.76 &  6017 & -14.10 &  -0.99 &  21.11 &   0.64 &  15.95 &  0.43 & 19.58 &   2.69 &   0.56 &  -0.11  \\
 4118 &		   &  18.31  &  1.56 &  5387 & -14.31 &  -8.88 &  22.82 &   5.22 &   1.10 &  0.42 & \nodata &  \nodata &  \nodata &   1.00  \\
 4129 &		   &  18.54  &  1.92 &  6088 & -14.60 &  -7.26 &  21.60 &   2.29 &   1.95 &  0.03 & \nodata &  \nodata &  \nodata &   1.00  \\
 4141 &		   &  18.71  &  1.73 &       & -14.74 &  -5.85 &  23.46 &   7.50 &   1.24 &  0.40 & \nodata &  \nodata &  \nodata &  -1.00  \\
 4175 &		   &  18.65  &  1.85 &  4628 & -15.26 &  -8.02 &  20.16 &   0.87 &   0.92 &  0.33 & 21.12 &   3.21 &   0.45 &   0.28  \\
 4200 &	    D182   &  16.84  &  1.72 &  5638 & -15.63 &   4.75 &  20.19 &   2.36 &   2.75 &  0.01 & \nodata &  \nodata &  \nodata &   1.00  \\
 4205 &		   &  19.45  &  2.60 & 62822 & -15.72 &  -7.05 &  20.92 &   1.57 &   3.59 &  0.41 & \nodata &  \nodata &  \nodata &   1.00  \\
\end{longtable}
}
\begin{table}[htb] 
\caption{Ellipticities} 
\begin{center}
\begin{tabular}{clcccc} 
\hline 
B/T     &num    &       & \hspace{-1.0cm}Bulge  &   & \hspace{-1.2cm}Disc \\
        &       &  $\bar \epsilon$ & $\sigma_\epsilon$ & 
	$\bar \epsilon$ &
	$\sigma_\epsilon$ \\
\\ 
\hline 
1	&54	&	0.21 & 0.14&	 \nodata&	\nodata\\ 
0.5-1	&14  	& 	0.21 & 0.16&	0.53 & 0.24\\ 
$<0.5$ &	61   & 0.21 & 0.15&	0.34 & 0.20\\
\end{tabular} 
\end{center} 
\label{elip}
\end{table}

{\footnotesize 
\begin{table}[htb]
\caption{Velocities} 
\begin {center} 
\begin{tabular}{lrrc}
Type & Number & $\bar{v}$ & $\sigma_v$  \\ 
% & & (km$\cdot$ s$^{-1}$) & (km$\cdot$ s$^{-1}$)\\
 \hline  \\
All			              & 162 & 6900 & 1182$\pm  66$ \\ 
\hspace {0.1cm}(B/T$\le$0.5)          &  61 & 6994 & 1137$\pm 103$ \\ 
\hspace {0.1cm} (B/T$>$0.5)           &  68 & 6904 & 1183$\pm 102$ \\ 
\hspace {0.3cm}E (B/T=1)              &  54 & 6864 & 1234$\pm 119$ \\ 
\hline
\hspace {0.5cm}E ($M_B < -17.5$) & 15 & 6871 & 1129$\pm 210$ \\
\hspace {0.5cm}E ($M_B > -17.5$) & 39 & 6862 & 1273$\pm 145$ \\
% el criterio para la separacion entre E gigantes y enanas es mB=17.35
\\
\hspace {0.5cm}E $(n> 2)$                 & 29 & 6825 & 1035$\pm 137$ \\ 
\hspace {0.5cm}E $(n< 2)$                 & 25 & 6910 & 1430$\pm 204$ \\ 
\hline
\hspace {0.3cm}S0   (0.5$<$B/T$<$1 )     & 14 & 7057 &   942$\pm 181$ \\  
\end{tabular} 
\end{center} 
\label{velo}
\end{table}
}

\clearpage

\clearpage
\newpage

\clearpage
\newpage

\begin{figure}  
\plotone{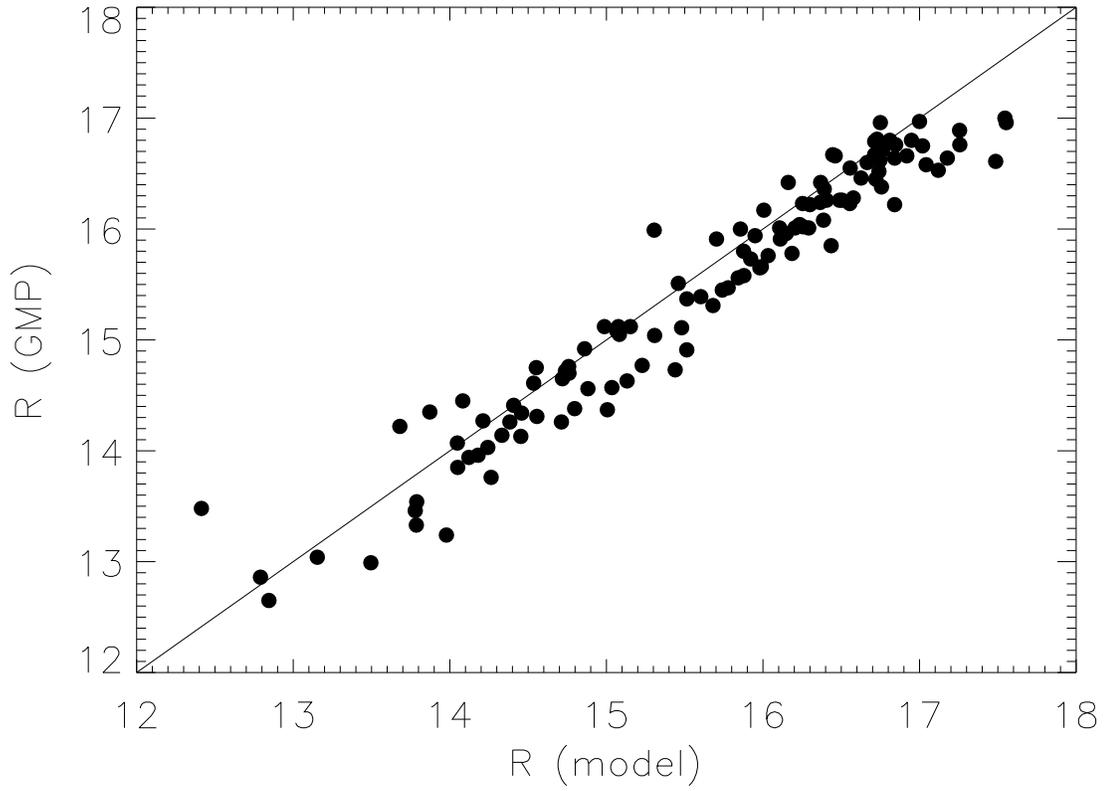}
\caption{Comparison between the $R$-band magnitudes quoted in the GMP catalog
and the galaxy magnitudes derived from our (Sloan-$r$ passband)
structural parameters.
\label{gmp}}
\end{figure}

\clearpage
\newpage

\clearpage
\newpage

\clearpage
\newpage

\clearpage
\newpage

\clearpage
\newpage

\clearpage
\newpage

\clearpage

\clearpage

\begin{thebibliography}{} 
\bibitem[]{} Adami, C., Mazure, A., Ulmer, M. P., \& Savine, C. 2001, A\&A, 
             371, 11
\bibitem[]{} Aguerri, J. A. L., \& Trujillo, I., 2002, MNRAS, 333, 633
\bibitem[]{} Aguerri, J. A. L., Iglesias-P\'aramo, J., V\'\i lchez, J. M. \& 
             Mu\~noz-Tu\~n\'on, C., 2003, in preparation
\bibitem[]{} Andredakis, Y. C., Peletier,  R. F., \& Balcells, M. 1995, 
             MNRAS, 275, 874 1995
\bibitem[]{} Andreon, S. 1996, A\&A, 314, 763
\bibitem[]{} Andreon, S., Davoust, E., \& Poulain, T. 1997, A\&AS, 126, 67
\bibitem[]{} Andreon, S., Davoust, E., Michard, R., Nieto, J.-L., \& 
             Poulain, T. 1995, A\&AS, 116, 429
\bibitem[]{} Andreon, S., \& Davoust, E. 1997, A\&A, 319, 747
\bibitem[]{} Arnouts, S., de Lapparent, V., Mathez, G., Mazure, A., 
             Mellier, Y., Bertin, E. \& Kruszewski, A.   1997
\bibitem[]{} Baier, F. W., Fritze, K., \& Tiersch, H. 1990, AN, 311, 89
\bibitem[]{} Balcells, M., Graham, A. W., Dom\'\i nguez-Palmero, L., \& 
             Peletier, R. F. 2003, ApJ, 582L, 79
\bibitem[]{} Balogh, M. L., Navarro, J. F., \& Morris, S. L.  
             2000, ApJ, 540, 113
\bibitem[]{} Beijersbergen, M., Hoekstra, H., van Dokkum, P. G., \& 
             van der Hulst, T. 2002, MNRAS, 329, 385
\bibitem[]{} Bernstein, G. M., Nichol, R. C., Tyson, J. A., Ulmer, M. P., \& 
             Wittmann, D. 1995, AJ, 110, 1507
\bibitem[]{} Binggeli, B., \& Jerjen, H. 1998, A\&A, 333, 17
\bibitem[]{} Binney, J. \& Tremaine, S. 1987, Galactic Dynamics (Princeton: 
             Princeton Univ. Press)
\bibitem[]{} Briel, U. G. 2001 et al.\ 2001, A\&A, 365, L60
\bibitem[]{} Burns, J. O., Roettiger, K., Ledlow, M., \& Klypin, A.  
             1994, ApJ, 427, L87
\bibitem[]{} Caldwell, N. 1987, AJ, 94, 1116
\bibitem[]{} Caon, N., Capaccioli, M., \& D'Onofrio, M. 1993, MNRAS, 265, 1013
\bibitem[]{} Castander, F. J.  et al.\ 2001, AJ, 121, 5
\bibitem[]{} Cellone, S. A., Forte, J. C., \& Geisler, D. 1994, AJSS, 93, 397
\bibitem[]{} Colless, M., \& Dunn, A. M. 1996, ApJ, 458, 435
\bibitem[]{} Conselice, C. J., \& Gallagher, J. S. III 1998, MNRAS, 297, 34
\bibitem[]{} Conselice, C. J.,  Gallagher, J. S. III, \& Wyse, R. F. G. 
             2001, ApJ, 559, 791
\bibitem[]{} Cote, S., Freeman, K. C., Carignan, C., \& Quinn, P. J. 
             1997, AJ, 114, 1313
\bibitem[]{} de Jong, R. S. 1996, A\&AS, 118, 557
\bibitem[]{} D'Onofrio, M., Capaccioli, M., \& Caon, N. 1994, MNRAS, 271, 523 
\bibitem[]{} Dressler, A. 1980, ApJSS, 42, 565 
\bibitem[]{} Drinkwater, M. J., Gregg, M. J., \& Colless, M. 
             2001, ApJ, 548, L39
\bibitem[]{} Edwards, S. A., Colless, M., Bridges, T. J., Carter, D., 
             Mobasher, B., \& Poggianti, B. M. 2002, ApJ, 567, 178
\bibitem[]{} Fitchett, M., \& Webster, R. 1987, ApJ, 317, 653
\bibitem[]{} Gallagher, J. S., Conselice, C. J., \& Wise, R. F. G. 
             2001, in ''Dwarf Galaxies and their environment''. 
	     Edited by Klaas S. De Boer, Ralf-Juergen Dettmar, and Uli Klein. 
	     ISBN 3826592646, p. 213
\bibitem[]{} Gavazzi, G., Zibetti, S., Boselli, A., Franzetti, P., 
             Scodeggio, M., \& Martocchi, S.  2001, A\&A, 372, 29
\bibitem[]{} Gerbal, D., Lima-Neto, G. B., Marquez, I., \& Verhagen, H. 
             1997, MNRAS, 285, L41
\bibitem[]{} Gnedin, O., 2003, ApJ, in press
\bibitem[]{} Godwin, J. G., Metcalfe, N., \& Peach, J. V. 1983, MNRAS, 202, 113
\bibitem[]{} Graham, A. W. 2001, AJ, 121, 820
\bibitem[]{} Graham, A. W. 2003, AJ, 125, 3398
\bibitem[]{} Graham, A.W., \& Colless, M.M.\ 1997, MNRAS, 287, 221 
\bibitem[]{} Graham, A., Guzm\'an, R. 2003, AJ, 125, 2936
\bibitem[]{} Graham, A., Lauer, T. R., Colless, M., \& Postman, M. 
             1996, ApJ, 465, 534
\bibitem[]{} Gurzadyan, V. G., \& Mazure, A. 2001, New astronomy, 6, 43
\bibitem[]{} Held, V. E., \& Mould, J. R. 1994, AJ, 107, 1307
\bibitem[]{} Jorgensen, I., \& Franx, M.  1994, ApJ, 433, 553
\bibitem[]{} Karachentsev, I. D., Karachentsev, V. E., Richter, G. M., 
             \& Vennik, J. A. 1995, A\&A, 296, 643
\bibitem[]{} Kashikawa, N.  et al.\ 1998, ApJ, 500, 750
\bibitem[]{} Kashikawa, N., Shimasaku, K., Yagi, M., Yasuda, N., Doi, M., 
             \& Okamura, S. 1995, ApJ, 452, L99
\bibitem[]{} Khosroshahi, H. G., Wadadekar, Y., Kembhavi, A., \& 
             Mobasher, B. 2000, ApJ, 531, L103
\bibitem[]{} Komiyama, Y. et al.\ 2002, ApJS, 138, 265
\bibitem[]{} Lahav, O., et al.\ 1995, Sci, 267, 859
\bibitem[]{} Lindner, U., Einasto, J., Einasto, M., Freudling, W., 
             Fricke, K., \& Tago, E.  1995, A\&A, 301, 329
\bibitem[]{} Lobo, C., Biviano, A., Durret, F., Gerbal, D., Le Fevre, O., 
             Mazure, A., \& Slezak, E. 1997, A\&ASS, 122, 409
\bibitem[]{} Lucey, J. R., Guzm\'an, R., Carter, D., \& Terlevich, R. J.  
             1991, MNRAS, 253, 584
\bibitem[]{} MacArthur, L.,A. Courteau, S., \& Holtzman, J. A.
             2003, ApJ, 582, 689
\bibitem[]{} Mar\'\i n-Franch, A., \& Aparicio, A. 2002, ApJ, 568, L74
\bibitem[]{} Mehlert, D., Saglia, R. P>, Bender, R., \& Wegner, G. 2000, 
             A\&AS, 141, 449
\bibitem[]{} Mobasher, B. et al.\ 2001, ApJS, 137, 279
\bibitem[]{} Mobasher, B., Guzm\'an, R., Aragon-Salamanca, A., \& Zepf, S. 
             1999, MNRAS, 304, 225
\bibitem[]{} Moore, B., Lake, G., Quinn, T., \& Stadel, J., 
             1999, MNRAS, 304, 465
\bibitem[]{} Neumann, D. M. et al.\ 2001, AA, 365, L74
\bibitem[]{} Pahre, M.  1999, ApJSS, 124, 127
\bibitem[]{} Quintana, H. 1979, AJ, 84, 15
\bibitem[]{} Rood, H. J., \& Baum, W. A. 1967, AJ, 72, 398
\bibitem[]{} Schindler, S., Binggeli, B., \& Bohringer, H. 1999, A\&A, 343, 420
\bibitem[]{} Secker, J.  Harris, W. E. 1997, PASP, 109, 1364
\bibitem[]{} S\'ersic, J. L.  1968, Atlas de galaxias australes, 
             C\'ordoba, Argentina: Observatorio Astron\'omico
\bibitem[]{} Terlevich, A. I., Caldwell, N., \& Bower, R. G. 
             2001, MNRAS, 326, 1547
\bibitem[]{} Trentham, N. 1998, MNRAS, 293, 71
\bibitem[]{} Trujillo, I., Aguerri, J. A. L., Guti\'errez, C. M., \&
             Cepa, J. 2001a, AJ, 122, 38 (T01A)
\bibitem[]{} Trujillo, I., Aguerri, J. A. L., Cepa, J., \& 
             Guti\'errez, C. M. 2001c,  MNRAS, 328, 977
\bibitem[]{} Trujillo, I., Aguerri, J. A. L., Cepa, J., \& 
             Guti\'errez, C. M. 2001b, MNRAS, 321, 269
\bibitem[]{} Trujillo, I., Aguerri, J. A. L., Guti\'errez, C. M.,
             Caon, N. \& Cepa, J. 2002a, ApJ, 573, L9 (T02A)
\bibitem[]{} Trujillo, I., Asensio Ramos, A., Rubino-Martin, J. A., Graham, A.
             W., Aguerri, J. A. L., Cepa, J., \& Gutierrez, C. M. 
	     2002b, MNRAS, 333, 510
\bibitem[]{} White, S., Briel, U. G., \& Henry, J. P. 1993, MNRAS, 261, L8
\bibitem[]{} Watanabe, M., Yamashita, K., Furuzawa, A., Kunieda, H., 
             Tawara, Y., \& Honda, H. 1999, ApJ, 527, 80
\bibitem[]{} Young, C. K., \& Currie, M. J. 1994, MNRAS, 268, L11
\bibitem[]{} Young, C. K., \& Currie, M. J. 1995, MNRAS, 273, 1141
\bibitem[]{} Zabludoff, A. I., \& Franx, M.  1993, AJ, 106, 1314
\end{thebibliography}
\end{document}